\documentclass[cernpreprint, atlasdraft=false, texlive=2016, UKenglish, texmf]{atlasdoc}
 
\usepackage{atlaspackage}
\usepackage{atlasbiblatex}
 
\usepackage{atlasphysics}
\usepackage{atlasheavyion}
\usepackage{atlasprocess}
\usepackage{atlasparticle}
\usepackage{atlasjetetmiss}
 
\graphicspath{{logos/}{figures/}}
 
\usepackage{ANA-HION-2019-58-PAPER-defs}
\usepackage{multirow}

\AtlasTitle{Measurement of the nuclear modification factor for muons from charm and bottom hadrons in Pb+Pb collisions at 5.02~TeV with the ATLAS detector}

\AtlasAbstract{
Heavy-flavour hadron production provides information about the transport properties and microscopic structure of the quark--gluon plasma created in ultra-relativistic heavy-ion collisions.
A measurement of the muons from semileptonic decays of charm and bottom hadrons produced in Pb+Pb and $pp$ collisions at a nucleon--nucleon centre-of-mass energy of 5.02~TeV with the ATLAS detector at the Large Hadron Collider is presented.
The Pb+Pb data were collected in 2015 and 2018 with sampled integrated luminosities of $208~\mathrm{\mu b}^{-1}$ and $38~\mathrm{\mu b^{-1}}$, respectively, and $pp$ data with a sampled integrated luminosity of $1.17~\mathrm{pb}^{-1}$ were collected in 2017.
Muons from heavy-flavour semileptonic decays are separated from the light-flavour hadronic background using the momentum imbalance between the inner detector and muon spectrometer measurements, and muons originating from charm and bottom decays are further separated via the muon track's transverse impact parameter.
Differential yields in Pb+Pb collisions and differential cross sections in $pp$ collisions
for such muons
are measured as a function of muon transverse momentum from 4~GeV to 30~GeV in the absolute pseudorapidity interval $|\eta| < 2$.
Nuclear modification factors for charm and bottom muons are presented as a function of muon transverse momentum in intervals of Pb+Pb collision centrality.   The bottom muon results are the most precise measurement of b quark nuclear modification at low transverse momentum where reconstruction of B hadrons is challenging.
The measured nuclear modification factors quantify a significant suppression of the yields of
muons from decays of charm and bottom hadrons, with stronger effects for muons from charm hadron decays.
}

\author{The ATLAS Collaboration}
 
\AtlasRefCode{HION-2019-58}
 
\AtlasJournalRef{\PLB 829 (2022) 137077}
 
\AtlasDOI{10.1016/j.physletb.2022.137077}
 
\PreprintIdNumber{CERN-EP-2021-153}

\hypersetup{pdftitle={ATLAS document},pdfauthor={The ATLAS Collaboration}}
 
\begin{document}
 
\maketitle

\section{Introduction}

Quark--gluon plasma (QGP) is a state of matter in which the quarks and gluons are deconfined from colour-neutral hadronic states.     Ultra-relativistic collisions of large nuclei create nuclei-sized droplets of QGP at temperatures in excess of 300--500~\MeV~\cite{Heinz:2013th,Romatschke:2017ejr}.   These droplets exist for a mere
$10^{-23}$ seconds and hence there is no way to fire an external probe at the droplet to investigate its properties.   Instead, the probes must be generated in the collision itself and then interact with the droplet.
Interactions with the QGP, both radiative and collisional, may provide key information regarding the properties and constituents of the QGP~\cite{Cao:2018ews}.
Specifically, the balance of radiative and collisional energy transfer depends on the mass of the constituents~\cite{Younus:2013rja,ColemanSmith:2012xb}.
Heavy quarks, charm and bottom, have masses much larger than the droplet temperature. Thus, they are produced in the initial collision via high-momentum-transfer interactions between incident quarks and gluons.  The strong-force interactions conserve the quantum numbers associated with the charm and bottom quarks. Thus, once created, these quarks can have substantial modifications to their momentum distributions when traversing the QGP, but they cannot be destroyed.
In addition, radiative energy loss is suppressed for heavy quarks by the so-called `dead-cone effect'~\cite{Dokshitzer:2001zm},
i.e.\ gluon radiation is suppressed at angles smaller than the quark's mass to energy ratio.
A key to constraining the relative contribution of radiative energy loss is to measure the modification of the momentum distributions for charm and bottom quarks separately, since the dead-cone effect will be more pronounced for bottom quarks than for charm quarks at the same momentum.
 
There are numerous publications detailing the modifications of momentum distributions of heavy-flavour hadrons measured in heavy-ion collisions via direct reconstruction and via decay leptons in heavy-ion collisions at the Relativistic Heavy Ion Collider (RHIC)~\cite{doi:10.1146/annurev.nucl.52.050102.090650} and the Large Hadron Collider (LHC)~\cite{Evans:2008zzb} -- current measurements and theory calculations are reviewed in Refs.~\cite{Dong:2019byy,Averbeck:2013oga}.
In nucleus--nucleus (A+A) collisions, each event is described by its centrality, which reflects the overlap of the colliding nuclei. The geometry of each event is calculated using a Monte Carlo (MC) Glauber model -- for details see Ref.~\cite{Glauber}.
The modification to particle yields in A+A collisions relative to \pp collisions is quantified by a nuclear modification factor \RAA defined as:
\begin{equation}
\RAA = \frac{N_\mathrm{AA} / N_\mathrm{evt}}{\langle \TAA \rangle  \times \sigma_{\pp}},
\label{eqn:raa}
\end{equation}
where $N_\mathrm{AA}$ is the number of observed particles of interest in \PbPb collisions, $N_\mathrm{evt}$ is the number of minimum-bias \PbPb events, $\langle \TAA \rangle$ is the average value of the nuclear thickness function,
and $\sigma_{\pp}$ is the particle production cross section in \pp collisions at the same collision energy.
If \RAA equals unity, the production in A+A collisions is the same as in \pp collisions but scaled up by the larger parton--parton luminosity, while $\RAA < 1$ indicates a suppression.
 
The nuclear modification factors for inclusive `heavy-flavour muons' (mostly muons from $D$ and $B$ meson semileptonic decays) have been measured in \PbPb collisions at $\sqn = 2.76~\TeV$ by the ATLAS experiment~\cite{HION-2015-06} and the ALICE experiment~\cite{Acharya:2020xzh}.
The ALICE experiment measured the nuclear modification factor for inclusive heavy-flavour electrons at the same energy~\cite{Acharya:2018upq,Adam:2016khe} and also at $\sqn = 5.02~\TeV$~\cite{Acharya:2019mom}.
Nuclear modification factors for charm hadrons $D^{0}$, $D_{s}$, $D^{*}$, and $\Lambda_{c}$ have been measured by the CMS~\cite{CMS-HIN-16-001} and ALICE~\cite{Acharya:2018hre,Acharya:2018ckj} experiments in \PbPb collisions at $\sqn = 5.02~\TeV$, and by the STAR~\cite{Adam:2018inb,Adamczyk:2014uip,STAR:2021tte,STAR:2019ank} experiment in Au+Au collisions at a lower energy of $\sqn = 200~\GeV$.
These measurements
indicate significant suppression for heavy-flavour hadrons in \PbPb collisions.
At transverse momentum (\pt) greater than $4~\GeV$, the prompt-$D^{0}$ \RAA is found to be consistent within the uncertainties with the \RAA of inclusive charged particles (dominated by $\pi$),
while at lower \pt values $D^{0}$ mesons have a smaller suppression (larger \RAA) compared to charged particles, as expected theoretically from the dead-cone effect.
 
The measurement presented here follows the previous ATLAS measurement of muons originating from heavy-flavour hadron decays in \PbPb collisions at $\sqn = 2.76~\TeV$~\cite{HION-2015-06}.
This Letter, based on the higher number of events in the combined 2015 and 2018 Pb+Pb datasets and the 2017 \pp dataset at $5.02~\TeV$, extends results to the transverse momentum range $4 < \pt < 30~\GeV$, and more importantly, further separates inclusive heavy-flavour muons into into contributions from charm hadron decays (charm muons) and bottom hadron decays (bottom muons).
The charm and bottom contribution separation is based on the muon track's transverse impact parameter, similar to the method used in previous ATLAS measurements~\cite{HION-2018-23, HION-2019-11}.
Results for charm and bottom muon cross sections in \pp collisions are shown as a function of muon \pt and compared with perturbative QCD calculations. The nuclear modification factor is presented as a function of muon \pt in various \PbPb centrality intervals.
Finally, the \RAA measurements, along with a previous measurement of the heavy-flavour muon momentum anisotropies~\cite{HION-2019-11}, are compared with the expectations from theoretical calculations.
The \RAA value quantifies the average energy loss, and the azimuthal anisotropy, $v_2$~\cite{HION-2019-11}, at high \pT, quantifies the azimuthal angle dependence of energy loss.
Simultaneous constraints on \RAA and $v_2$ for the same final state are important in distinguishing the relative impacts of different heavy-quark energy loss mechanisms and in understanding the influence of the initial QGP droplet geometry on the resulting evolution of kinematics of heavy-flavour quarks in the medium.

\section{ATLAS detector}
 
\newcommand{\AtlasCoordFootnote}{
ATLAS uses a right-handed coordinate system with its origin at the nominal interaction point (IP)
in the centre of the detector and the \(z\)-axis along the beam pipe.
The \(x\)-axis points from the IP to the centre of the LHC ring,
and the \(y\)-axis points upwards.
Cylindrical coordinates \((r,\phi)\) are used in the transverse plane,
\(\phi\) being the azimuthal angle around the \(z\)-axis.
The pseudorapidity is defined in terms of the polar angle \(\theta\) as \(\eta = -\ln \tan(\theta/2)\).
Angular distance is measured in units of \(\Delta R \equiv \sqrt{(\Delta\eta)^{2} + (\Delta\phi)^{2}}\).}

The ATLAS detector~\cite{PERF-2007-01,ATLAS-TDR-2010-19,PIX-2018-001} at the LHC covers nearly the entire solid angle around the collision point.\footnote{\AtlasCoordFootnote}
It consists of an inner tracking detector surrounded by a thin superconducting solenoid, electromagnetic and hadronic calorimeters,
and a muon spectrometer incorporating three large superconducting toroidal magnets with eight coils each.
The inner-detector system (ID) is immersed in a \SI{2}{\tesla} axial magnetic field
and provides charged-particle tracking in the range \(|\eta| < 2.5\).
 
The high-granularity silicon pixel detector covers the vertex region and typically provides four measurements per track, with
the first hit typically in the insertable B-layer installed before Run~2~\cite{ATLAS-TDR-2010-19,PIX-2018-001}.
It is followed by the silicon microstrip tracker which usually provides eight measurements per track.
These silicon detectors are complemented by the transition radiation tracker (TRT),
which enables radially extended track reconstruction up to \(|\eta| = 2.0\).

The calorimeter system covers the pseudorapidity range \(|\eta| < 4.9\).
Within the region \(|\eta|< 3.2\), electromagnetic calorimetry is provided by barrel and
endcap high-granularity lead/liquid-argon (LAr) calorimeters,
with an additional thin LAr presampler covering \(|\eta| < 1.8\)
to correct for energy loss in material upstream of the calorimeters.
Hadronic calorimetry is provided by the steel/scintillator-tile calorimeter,
segmented into three barrel structures within \(|\eta| < 1.7\), and two copper/LAr hadronic endcap calorimeters.
The solid angle coverage is completed with forward copper/LAr and tungsten/LAr calorimeter modules (FCal),
covering the forward regions of $3.1 < |\eta| < 4.9$,
optimized for electromagnetic and hadronic measurements respectively.
The minimum-bias trigger scintillators detect charged
particles over $2.07<|\eta|<3.86$ using two hodoscopes of 12 counters positioned at $z=\pm$3.6~m.
The zero-degree calorimeters (ZDC) measure neutral particles at pseudorapidities $|\eta| \geq 8.3$
and consist of layers of alternating quartz rods and tungsten plates.
 
The muon spectrometer (MS) comprises separate trigger and
high-precision tracking chambers measuring the deflection of muons in a magnetic field generated by superconducting air-core toroids.
The field integral of the toroids ranges between \num{2.0} and \SI{6.0}{\tesla\metre}
across most of the detector.
A set of precision chambers covers the region \(|\eta| < 2.7\) with three layers of monitored drift tubes,
complemented by cathode-strip chambers in the forward region, where the background is highest.
The muon trigger system covers the range \(|\eta| < 2.4\) with resistive-plate chambers in the barrel, and thin-gap chambers in the endcap regions.
Events of interest are selected to be recorded by the first-level trigger (L1) system implemented in custom hardware,
followed by selections made by algorithms implemented in software in the high-level trigger (HLT)~\cite{TRIG-2016-01}.
The first-level trigger selects events from the \SI{40}{\MHz} bunch crossings at a rate below \SI{100}{\kHz} (\SI{75}{\kHz}) for \pp (\PbPb) collisions,
and the high-level trigger reduces the average event output rate to about \SI{1.2}{\kHz} for recording.
An extensive software suite~\cite{ATL-SOFT-PUB-2021-001} is used for real and simulated data reconstruction and analysis, for operation and in the trigger and data acquisition systems of the experiment.

\section{Event selection}
 
The \pp data used in this analysis were recorded with the ATLAS detector in 2017,
while the \PbPb data were recorded in 2015 and 2018.
Both the \pp and \PbPb events were selected online using a trigger that requires a muon at the L1 and HLT with a \pt larger than $4~\GeV$~\cite{TRIG-2016-01, TRIG-2018-01}.
After selecting run periods when the detector subsystems were operational and taking into account
the fraction of the total luminosity sampled by the triggers,
the datasets used in the analysis correspond to integrated luminosities
of $1.17~\ipb$ for \pp data, $208~\mathrm{\mu b}^{-1}$ for 2015 \PbPb data, and $38~\mathrm{\mu b}^{-1}$ for 2018 \PbPb data.
The 2017 \pp data, with a small average
number of interactions per bunch crossing in the range from $0.4$ to 4,
were collected to serve as the baseline for \PbPb collision measurements at the same centre-of-mass energy per nucleon pair.
Only a small fraction of events from the 2018 Pb+Pb data was recorded by the low \pT muon trigger used in this analysis. Other muon triggers in the 2018 Pb+Pb data were studied and improved statistical uncertainties while increasing systematic uncertainties, and are thus not incorporated.   The 2015 and 2018 results are consistent and thus combined in these results.
The selected \PbPb events are further required to satisfy offline minimum-bias \PbPb collision criteria,
identical to those used in Ref.~\cite{HION-2019-11}.
This additional requirement identifies and rejects $0.2\%$ of the selected events as pile-up events, based on a combination of the total transverse energy measured in the FCal, denoted by \sumETFcal, and the ZDC energy.

The centrality of each \PbPb event is characterized by its \sumETFcal value.  For the results shown here, the minimum-bias \sumETFcal distribution is divided into percentiles ordered from the most central (large \sumETFcal, small impact parameter) to the most peripheral (small \sumETFcal, large impact parameter):  0--10\%, 10--20\%, 20--30\%, 30--40\%, and 40--60\%.
The interval 0--100\% corresponds to the total \PbPb inelastic cross section~\cite{guillaud:in2p3-00021835}.
An MC Glauber~\cite{Glauber} model
is used to calculate $\langle \TAA \rangle$ for each centrality interval~\cite{HION-2016-06}.

Muons with $4 < \pt < 30~\GeV$ and $|\eta|<2$ reconstructed in both the ID and the MS are selected and required to pass `medium' selection requirements, detailed in Ref.~\cite{PERF-2015-10}.
Selected muons are required to be matched with an online muon candidate that fires the event trigger.
Each muon is assigned a weight which is the inverse of the product of the reconstruction and trigger efficiencies, evaluated for muons as a function of their kinematic variables,
and in the case of \PbPb data as a function of centrality as well.

The muon reconstruction and identification efficiency is factorized as the product of the individual reconstruction efficiencies in the ID and MS, where the ID and MS matching efficiency is included in the MS efficiency.
The ID and MS efficiencies in \pp collisions are determined in the large \pp dataset collected in 2017 at $\sqrt{s} = 13~\TeV$, using the tag-and-probe method on \Jmm events as detailed in Ref.~\cite{PERF-2015-10}.
They are applied to the \pp data at $\sqrt{s} = 5.02~\TeV$ used in this analysis, in fine intervals of muon \pt and $\eta$.
The difference between the muon reconstruction efficiencies at $\sqrt{s} = 5.02~\TeV$ and $\sqrt{s} = 13~\TeV$ due to different pile-up contributions and multiplicities is found to be negligible~\cite{PERF-2015-10}.
The MS efficiencies obtained from \pp and \PbPb data collected in the same year show no significant difference.
Thus, the MS efficiencies for 2015 and 2018 \PbPb data are obtained from the large \pp datasets at $\sqrt{s} = 13~\TeV$ in the corresponding years in fine intervals of muon \pt and $\eta$.
The resulting MS efficiency plateaus at 97\% at $\pt > 7~\GeV$.
The ID efficiency in \PbPb events is obtained from \PbPb data using the same \Jmm tag-and-probe method in intervals of muon \pt and $\eta$.
The measured \PbPb ID efficiency is about 98\%
with the efficiency in 2018
being 3\% lower than that in 2015 at $1 < |\eta| < 2$ independent of \pt, while no difference between 2015 and 2018 is observed at $|\eta| < 1$.
No centrality dependence is observed for MS and ID efficiencies for muons in \PbPb collisions.

The initial estimations of the muon trigger efficiency
are obtained from \Jmm \PYTHIA[8]~\cite{Sjostrand:2014zea} simulations in fine intervals of muon \pt and $\eta$ using the tag-and-probe method~\cite{TRIG-2018-01}.
All generated events were passed through a \GEANT simulation~\cite{geant,SOFT-2010-01} of the ATLAS detector under the same conditions as present during data-taking and were digitized and reconstructed in the same way as the data.
The same simulated trigger efficiency is used in \pp and \PbPb data. Relative efficiency differences between simulations and data and between \pp and \PbPb collisions are covered by additional corrections described as follows.
Mis-modelling of the trigger performance in simulation, quantified by the ratio of measured efficiencies in data and the simulations, is accounted for by applying a multiplicative correction not exceeding 10\%.
To account for the different online muon momentum scale in \pp and \PbPb data, as well as the slightly different trigger performance in 2015 and 2018,
an additional correction factor, determined by the \PbPb to \pp data-driven efficiency ratio, is applied to \PbPb data as a function of muon \pt and $\eta$, the centrality, and the year of \PbPb data-taking.

\section{Signal extraction}
 
\begin{figure}[hbt]
\centering
\includegraphics[width=0.45\linewidth]{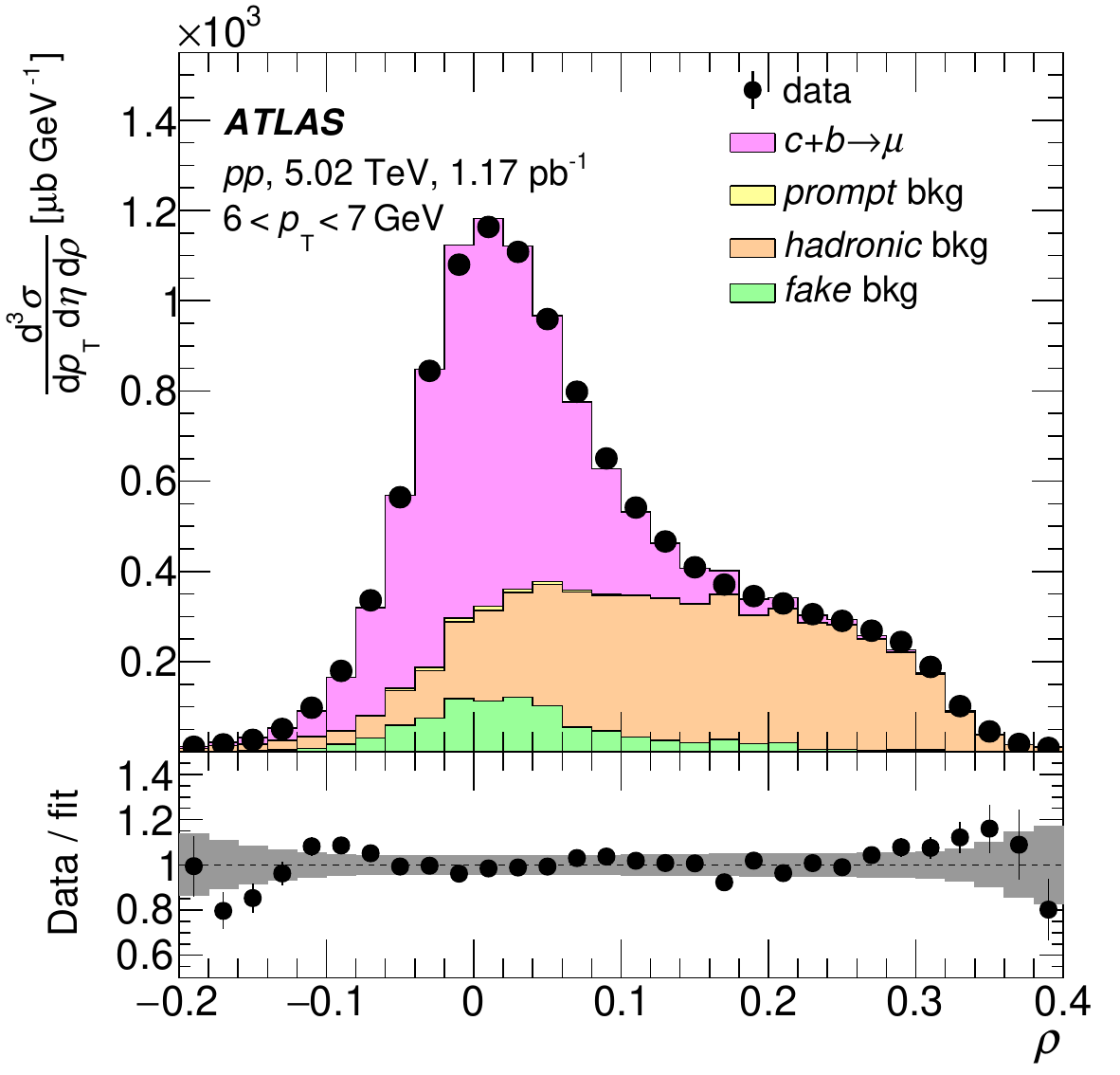}
\includegraphics[width=0.45\linewidth]{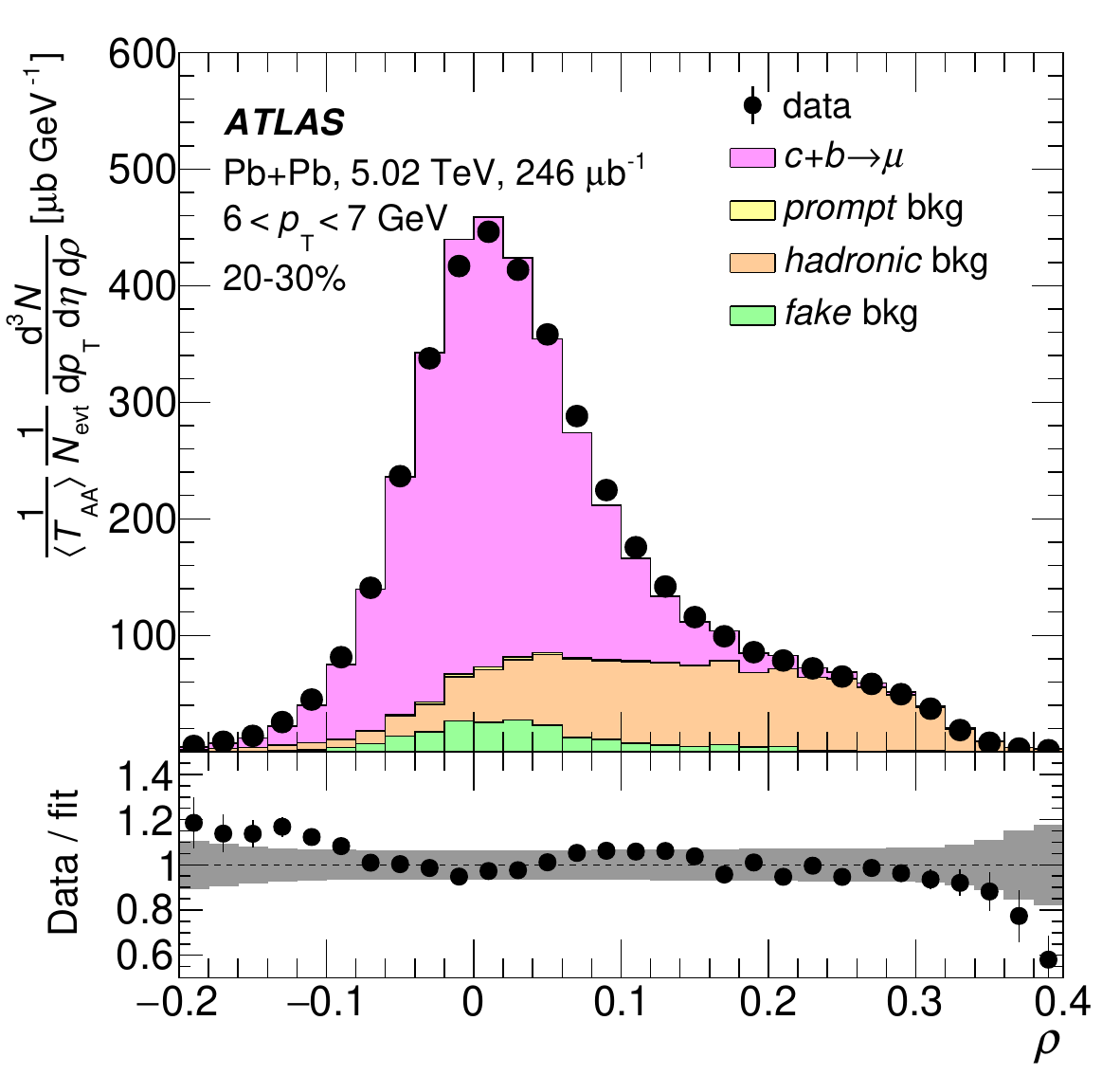}
\caption{\label{fig:fit_rho} Fit results of the $\rho$ distribution for muons with $6 < \pt < 7~\GeV$ in \pp collisions (left) and in \PbPb collisions with 20--30\% centrality (right). The ratios of the data to fit results are shown in the lower panels. The prompt-muon background contributions are very small in the fitted muon kinematic region.
The grey bands in the lower panels indicate the statistical and systematic uncertainties combined in quadrature.}
\end{figure}

\begin{figure}[hbt]
\centering
\includegraphics[width=0.45\linewidth]{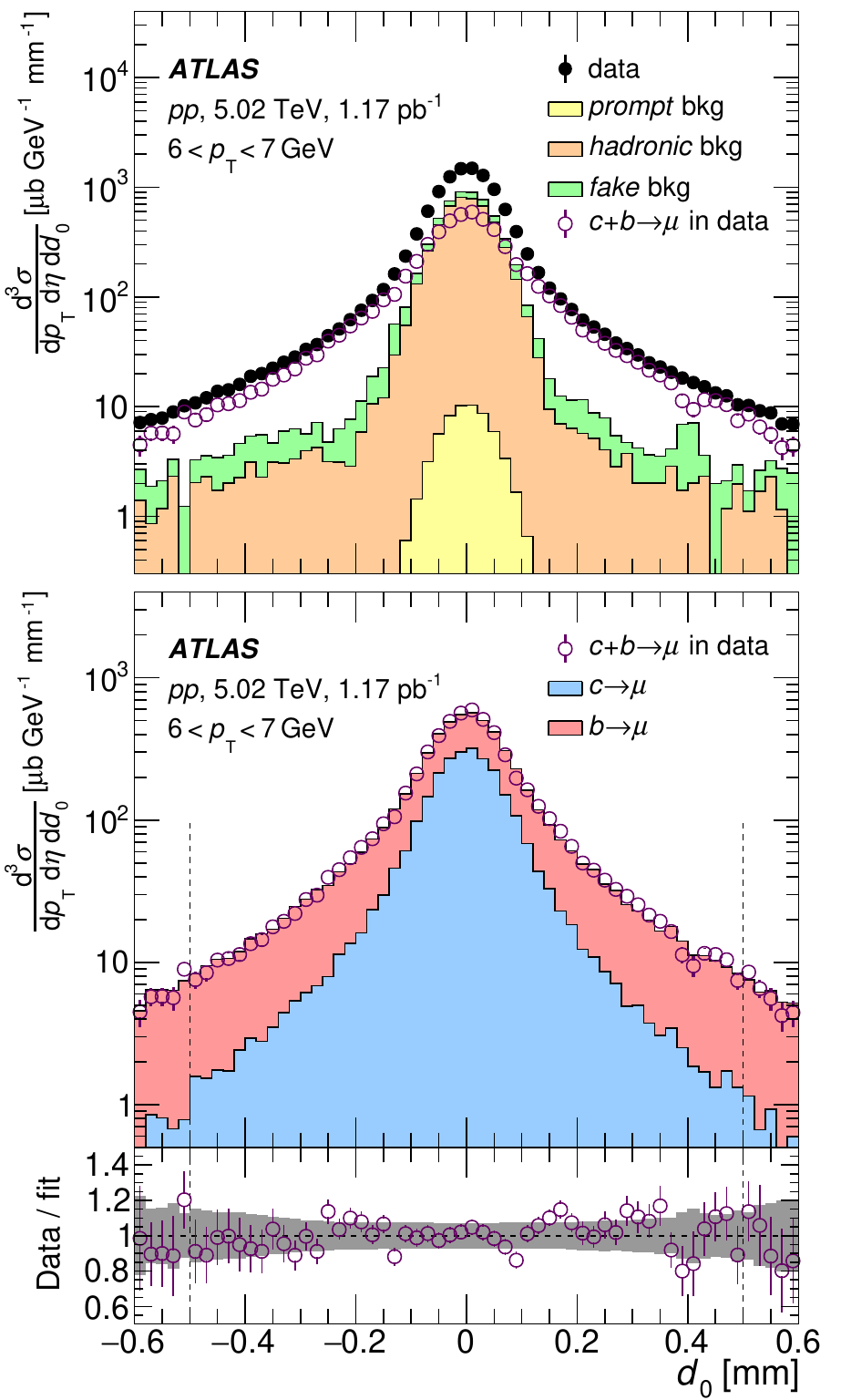}
\includegraphics[width=0.45\linewidth]{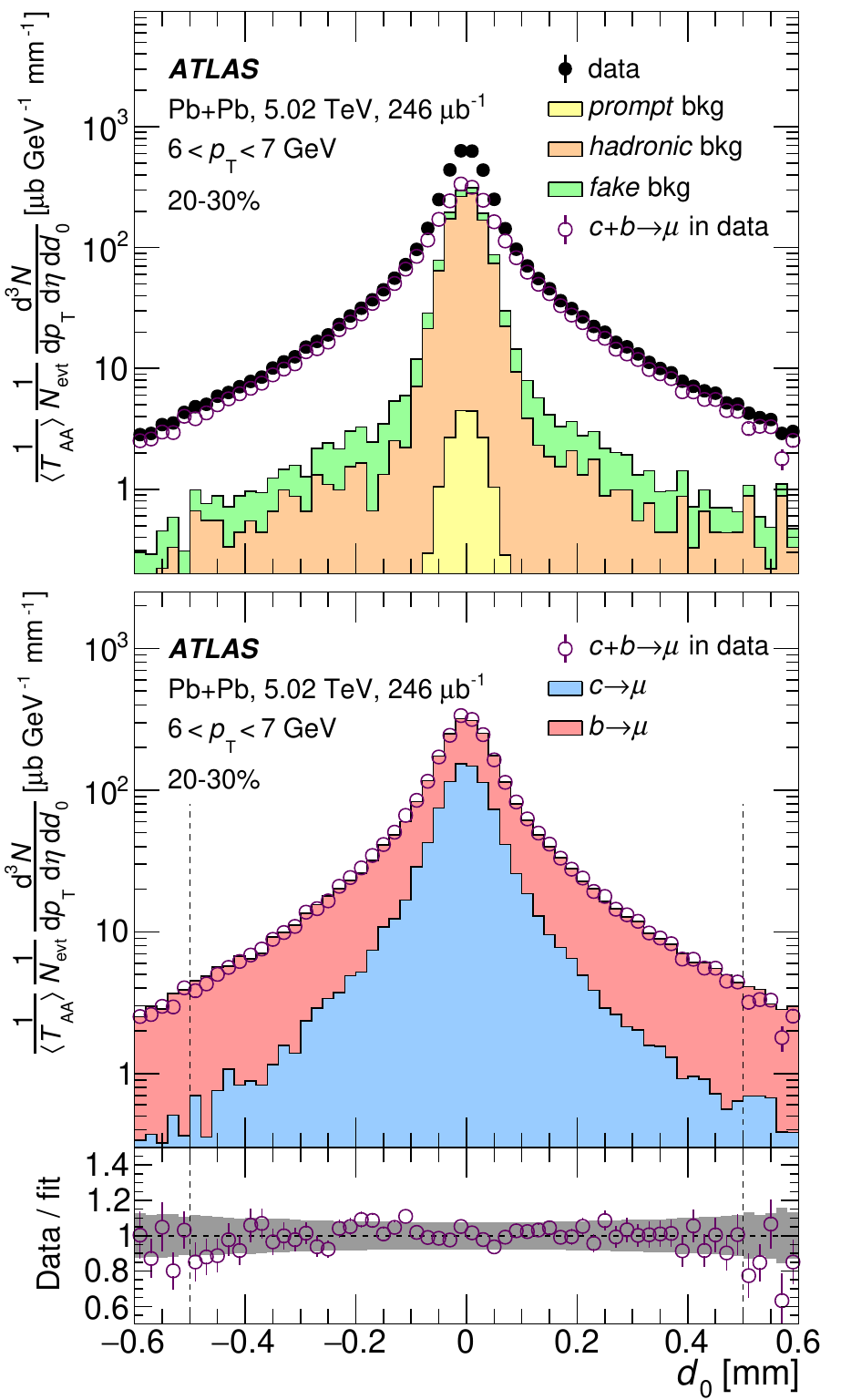}
\caption{\label{fig:fit_dca} (Top) \dcaT templates for different background contributions, and data \dcaT distributions before (solid points) and after (open points) background subtraction for muons with $6 < \pt < 7~\GeV$ in \pp collisions (left) and in \PbPb collisions with 20--30\% centrality (right).
(Bottom) Fit results of the background-subtracted \dcaT distribution
for muons with $6 < \pt < 7~\GeV$ in \pp collisions (left) and in \PbPb collisions with 20--30\% centrality (right).
The vertical dashed lines indicate the \dcaT range, $|\dcaT| < 0.5~\mathrm{mm}$, used in the fit procedure.
Data to fitted result ratios are shown in the bottom panels.
The grey bands indicate the statistical and systematic uncertainties combined in quadrature.
}
\end{figure}

As detailed in previous ATLAS publications~\cite{STDM-2011-07,HION-2015-06,HION-2018-23,HION-2019-11},
the background contributions in the selected muon samples (labelled `bkg' in Figures~\ref{fig:fit_rho} and~\ref{fig:fit_dca}) have three components.
The first one is called the `prompt-muon background' and includes contributions from decays of
non-open--heavy-flavour particles such as direct quarkonia, low-mass resonances, $\tau$-leptons, and massive electroweak $W/Z$ bosons.
The second component is the `hadronic background' resulting from $\pi$ and $K$ decaying into muons in the volume of the ID or punching through the calorimeter.
The last component is from random combinations of uncorrelated track segments from the ID and MS, called the `fake-muon background'.
The prompt-muon background, expected to be small compared to signal muons, is estimated in simulations.   These simulations are scaled to match existing measurements of different background sources under consistent fiducial selections.
The main prompt-muon background for muons with $20 < \pT < 30~\GeV$ is from $W$ decays, and is approximately 3\% in \pp collisions at 10\% (5\%) in 0-10\% (40-60\%) \PbPb collisions based on previous ATLAS measurements~\cite{HION-2018-02, HION-2018-25}.
Hadronic and fake-muon backgrounds are subtracted from the signal muons by fitting the muon momentum imbalance, $\rho = \rhodef$, where $p^\mathrm{ID}$ is the muon momentum measured in the ID, and $p^\mathrm{MS}$ is that measured in the MS corrected for the energy loss inside the calorimeter. The $\rho$ distribution shapes of hadronic and fake-muon backgrounds are extracted from simulations while their yields are determined from the fit procedure.

The prompt-muon background contribution in \pp events is estimated from \PYTHIA[8] simulations of prompt \Jpsi, \psip, and \Ups{n} production based on a non-relativistic QCD colour-octet model~\cite{NRQCD:1995}, and
from $W$ and $Z$ production simulated with the \POWHEGBOX[v2] generator~\cite{Alioli:2010xd} interfaced to the \PYTHIA[8] parton shower model.
The CT10 PDF set~\cite{Lai:2010vv} was used in the matrix element, while the CTEQ6L1 PDF set~\cite{Pumplin:2002vw} was used for the modelling of non-perturbative effects in the initial-state parton shower.
The simulated prompt-muon background processes are all scaled by process-dependent single scaling factors to match previous ATLAS measurements in \pp collisions at $\sqrt{s} = 5.02~\TeV$~\cite{HION-2014-05, HION-2018-02}.
Other contributions from low-mass resonances and $\tau$-leptons in the measured muon \pt range are found to be less than $1\%$~\cite{STDM-2011-07, HION-2018-23} and are neglected in this analysis.
In \PbPb collisions, the estimated prompt-muon background rates take into account nuclear modifications measured in Refs.~\cite{HION-2016-07,CMS-HIN-16-023,HION-2018-25, HION-2018-10}.

Muons from hard scattering have a symmetric $\rho$ distribution peaked at zero,
while the hadronic background has a broader $\rho$ distribution and the peak shifted toward higher values.
This change in shape is due to differences between the actual energy loss of this background and the estimated energy loss based on muon simulations.
Only energy loss of muons is properly corrected for and added to the MS momentum measurement.
The different shapes of the $\rho$ distributions for the hadronic background and the other muons allow the hadronic background to be isolated using a template-fitting procedure~\cite{HION-2018-23,HION-2019-11}.
The yields of inclusive heavy-flavour muons and hadronic background muons are extracted from the $\rho$ template fit.
The templates for the charm and bottom muon $\rho$ distributions are determined from
multijet hard-scattering \PYTHIA[8] \pp collision events at $\sqrt{s} = 5.02~\TeV$ filtered for the presence of a generator-level muon produced with parameter values as in the A14 tune~\cite{ATL-PHYS-PUB-2014-021} and using the \NNPDF[23lo] parton distribution functions~\cite{Ball:2012cx}.
The templates for the charm and bottom muon $\rho$ distributions (and track transverse impact parameter, \dcaT, distributions, described below) in
multijet hard-scattering QCD \PYTHIA[8] samples are found to be identical to those from non-diffractive QCD \PYTHIA[8] samples, while hard-scattering QCD \PYTHIA[8] samples have much higher muon filter efficiency.
The $\rho$ templates for the
prompt-muon background are obtained from the simulation procedure
described in the previous paragraph.
The hadronic background and fake-muon $\rho$ templates are obtained from non-diffractive QCD simulations of \pp collisions at $\sqrt{s} =5.02~\TeV$ in \PYTHIA[8], with the A14 tune and \NNPDF[23lo] parton distribution functions.
The fake-muon contribution is fixed relative to the hadronic background, with the ratio obtained from simulations.
The momenta of the reconstructed muons selected in \PYTHIA[8] simulations are calibrated to match the muon momentum response in \pp and \PbPb data.
The calibration is performed via shift and smearing parameters~\cite{HION-2019-11} determined from the invariant mass response in \Jmm events.
The same calibration is applied to all muon candidates including those from background contributions.
Alternative calibrations for background contributions are used for the systematic uncertainty evaluation.
 
Examples of the $\rho$ template fit are shown in \Fig{\ref{fig:fit_rho}}
for muons with $6 < \pt < 7~\GeV$
in \pp collisions and 20--30\% centrality \PbPb collisions.
The signal muon $\rho$ distribution shape shows no obvious dependence on muon \pt, but is found to be broader in the more forward pseudorapidity region and more central \PbPb collisions, both due to poorer muon momentum resolution in the ID.
The hadronic background $\rho$ distribution becomes broader at higher \pt and in more central collisions.
 
Charm and bottom muons are further separated using the muon track's transverse impact parameter, \dcaT, which is calculated relative to the beam spot~\cite{PERF-2015-01}.
Due to the different lifetimes of charm and bottom hadrons
(approximately $1.5 \times 10^{-12}$~s for $B$ mesons, $1.0 \times 10^{-12}$~s for $D^+$ and $0.4 \times 10^{-12}$~s for $D^0$ mesons),
the corresponding muons have different \dcaT distributions, and their fractional contributions can be extracted using a template-fitting procedure.
Background contributions are subtracted from data distributions of \dcaT as shown in the upper panels in \Fig{\ref{fig:fit_dca}}.
For each of the three background sources, the \dcaT shape is determined in simulations.
In \pp data analysis, the \dcaT shape templates of various background sources are obtained from the \PYTHIA[8] simulations used to build the $\rho$ templates.
In \PbPb data analysis, the same \PYTHIA[8] events overlaid with minimum-bias \PbPb events collected in 2015 are used to build the background \dcaT templates to
approximate background distributions in \PbPb collisions.
The \dcaT distributions from high-quality prompt tracks in simulations and 2018 \PbPb data are smeared to match those in 2015 \PbPb data.
The prompt-muon background \dcaT distribution normalization is constrained by the yield estimates from MC simulations. The hadronic and fake-muon background \dcaT distribution normalization factors are extracted from the $\rho$ template fit.
The signal muon \dcaT  distribution in \PbPb data is narrower than in \pp data because of a smaller transverse beam size.
The background \dcaT distribution becomes moderately narrower with increasing \pt and shows no evident centrality dependence.
After subtraction, the remaining \dcaT distribution in data contains only contributions from heavy-flavour muons.

The \dcaT fit is performed in the range of $|\dcaT| < 0.5~\mathrm{mm}$ as
events with $|\dcaT| > 0.5~\mathrm{mm}$ are statistically limited in both data and simulations and have little sensitivity to the charm muon contribution.
For \pp data, the charm and bottom muon \dcaT templates are obtained from the muon-filtered multijet \PYTHIA[8] simulations at $\sqrt{s} =5.02~\TeV$, as done for $\rho$ templates.
Bottom muons contain the $b\rightarrow c\rightarrow\mu$ cascade contribution.
The \dcaT distributions of signal muons show no obvious dependence on the muon \pt,
but they become broader with increasing parent $B$ and $D$ meson \pt.
The simulated samples from \PYTHIA[8] are reweighted to match the inclusive $B$ and $D$ meson \pt spectra from fixed-order next-to-leading-log (FONLL) resummation calculations~\cite{Cacciari:1998it,Cacciari:2001td}.
The charm and bottom baryon-to-meson ratios in the simulations are corrected to match the measured values in Refs.~\cite{Lisovyi_2016, Aaltonen_2008}.  The baryon-to-meson ratio is treated the same in \pp and \PbPb, and allowed to vary independently for each system by a factor of two, based on values reported in Ref.~\cite{CMS:2019uws}, in determining systematic uncertainties.
The yield of $D^+$ relative to $D^0$ is corrected to match the measured value in Refs.~\cite{Zyla:2020zbs, lohrmann2011summary}.
In the \PbPb analysis, the charm and bottom muon \dcaT templates are obtained from the \pp \PYTHIA[8] simulations overlaid with minimum-bias \PbPb events from 2015, similar to what is done for the background \dcaT templates.
Besides the FONLL resummation and baryon-to-meson corrections as applied in \pp collisions, an additional correction is applied to match the modified charm and bottom hadron \pt spectra measured in \PbPb collisions by ALICE~\cite{Acharya:2018hre} and CMS~\cite{CMS-HIN-16-001}.
Examples of \dcaT template fits for muons with $6 < \pt < 7~\GeV$ are shown in \Fig{\ref{fig:fit_dca}} for \pp collisions and 20--30\% centrality \PbPb collisions.
The relative fractions of charm and bottom muons are extracted from the \dcaT template fit.
 
\section{Systematic uncertainties}

Systematic uncertainties associated with the various steps of the analysis are assessed.
The measured cross section in \pp collisions, per-event yields in \PbPb collisions and \RAA are recalculated
by systematically varying the effect of each source of uncertainty
and then compared with the nominal results. The resulting difference is assigned as a systematic uncertainty.
Each group of systematic uncertainties described in the following subsections are considered as uncorrelated,
and are therefore summed in quadrature.
Some sources of systematic uncertainties that are correlated between \pp and \PbPb collisions partially cancel out in \RAA.

\subsection{Muon correction uncertainties}
The systematic uncertainties from the muon reconstruction efficiency and muon trigger efficiency corrections for \pp collisions are dominated by the uncertainty in determining these efficiencies in data with the tag-and-probe method.
These are evaluated following the procedures in previous ATLAS measurements~\cite{PERF-2015-10, TRIG-2018-01}, including variations in the tag-and-probe efficiency extraction method, online--offline matching requirement, and muon purity in the selected sample.
The \PbPb efficiency is affected by the systematic uncertainty sources mentioned above, due to the use of the \pp efficiency in the factorized treatment.
An additional uncertainty in \PbPb is associated with the determination of the ID reconstruction efficiency in data using the tag-and-probe method.
Uncertainties in the trigger efficiency correction are determined from the \PbPb to \pp efficiency ratio and residual discrepancies between fully corrected \PbPb muon spectra measured in the 2015 and 2018 data-taking periods.
Apart from specific \PbPb uncertainties, other uncertainties are correlated between collision systems and thus cancel out in \RAA.
 
\subsection{Background removal uncertainties}
The systematic uncertainty in the momentum imbalance templates includes the effect of the uncertainty in the muon momentum calibration parameters,
the uncertainty due to the dedicated hadronic-background calibration, and the uncertainty due to $\rho$--$\dcaT$ correlation in the hadronic background.
Charm muon yields are more sensitive to the background removal procedure as charm muons have narrower \dcaT distributions around 0 where the background contamination is large.
The systematic uncertainties in the calibration parameters, which originate from the uncertainty in the determination of \Jmm invariant mass~\cite{BPHY-2012-02},
contribute $3\%$ relative uncertainty to the final charm muon yields
and $1\%$ to the bottom muon yields.
Since the hadronic background could have a different momentum scale than that for real muons, the results are also examined using a data-driven hadronic-background momentum calibration in the background-dominated region $\rho > 0.2$. The difference is included as a systematic uncertainty which is about $10\%$ ($4\%$) for charm (bottom) muon yields at low \pt and decreases with increasing \pt.
To account for the small $\rho$--$\dcaT$ correlation in hadronic background, the corresponding $\rho$ templates in different \dcaT selections are used and the resulting difference is assigned as an additional systematic uncertainty.
The relative uncertainty due to the background $\rho$--$\dcaT$ correlation is about $3\%$ for charm muon yields and less than $1\%$ for bottom muon yields.
The $\rho$-template-related systematic uncertainties are treated as correlated between \pp and \PbPb data, and partially cancel out in \RAA.

In the analysis, the size of each prompt-muon background contribution is held at a fixed value obtained from \PYTHIA[8] simulations scaled to match existing measurements.
To assess the associated uncertainty, the \pp and \PbPb analyses are repeated while varying the estimated sizes of the different prompt-muon contributions by the corresponding experimental uncertainties in their production rates~\cite{HION-2014-05, HION-2018-02} and \RAA values~\cite{HION-2016-07,CMS-HIN-16-023,HION-2018-25, HION-2018-10}.
The sensitivity to the fake-muon rate estimation is evaluated by varying the fake-muon candidate
definition in simulations with different truth-level and reconstruction-level objects matching criteria.
The resulting variation of the fake-muon rate is about 20--40\%, and produces $5\%$ variations, on average, in the measured charm and bottom yields. The fake-muon systematic uncertainty is assumed to be uncorrelated between the \pp and \PbPb results.

\subsection{Charm--bottom separation uncertainties}
The systematic uncertainties in the impact parameter template,
that are common for \pp and \PbPb results, include several components.
Uncertainties in the FONLL calculations are evaluated based on Ref.~\cite{Cacciari:1998it,Cacciari:2001td}.
Uncertainties coming from baryon-to-meson ratio mis-modelling correction use measurements published in Ref.~\cite{Lisovyi_2016, Aaltonen_2008}.
Uncertainties in $D^+/D^0$ mis-modelling correction are based on reported uncertainties in Ref.~\cite{Acharya:2021cqv,ALICE:2011aa}.
They are propagated to the \pp and \PbPb results,
and are treated as correlated between the \pp and \PbPb results.
The muon \pt spectra predicted by FONLL calculations are weighted to match the measured charm and bottom muon spectra reported in this Letter. The resulting weighting factors are used to adjust signal \dcaT templates to quantify the bias due to observed mis-modelling in the FONLL calculations.
The resulting difference of $1$--$5\%$ in the muon yields is assigned as an additional uncertainty and is treated as correlated between the \pp and \PbPb results.
Uncertainties in the nuclear modification factor for parent hadrons in \PbPb are propagated
using the experimental uncertainties in $D$ and $B$ meson \RAA values from Refs.~\cite{Acharya:2018hre,CMS-HIN-16-001}.
Minimum-bias \PbPb events collected in 2018, instead of 2015 events, are used to overlay with \PYTHIA[8] simulations to test the sensitivity to slightly different overlay conditions.
Resulting changes in the muon yields due to different overlay conditions are assigned as a systematic uncertainty in the \PbPb results.
Uncertainties in the determination of \dcaT shift and smearing parameters are found to have a negligible impact, less than $0.5\%$, on the extracted charm and bottom yields.
 
\subsection{Global normalization uncertainties}
In the 2017 \pp data, the LUCID-2 detector~\cite{LUCID2} is used for the primary luminosity measurement. The uncertainty in the integrated luminosity is derived using the methods described in Ref. \cite{ATLAS-CONF-2019-021}, and is $1.6\%$.
For \PbPb collisions, the systematic uncertainty in $\avgTAA$ is estimated by varying the MC Glauber model parameters as detailed in Ref.~\cite{HION-2016-06}.
 
\subsection{Uncertainty summary}
 
Table~\ref{tab:systematics_summary} summarizes relative uncertainties in the measurement of charm and bottom muon production in \pp and \PbPb collisions, and in the nuclear modification factor \RAA.
The leading sources of uncertainty in all muon \pt bins and \PbPb collision centralities are the background removal and charm--bottom separation uncertainties, and at low muon \pt in central Pb+Pb collisions the evaluation of the muon efficiency.
 
\begin{table}[thp]
\caption{\label{tab:systematics_summary}
Contributions to systematic uncertainties given in percent for the cross section in \pp, yields in \PbPb, and nuclear modification factor of charm and bottom muons. Ranges indicate the minimum and maximum systematic uncertainties found in all muon \pt bins and centralities for a given source.
}
\begin{center}
\renewcommand{\arraystretch}{1.4}
\resizebox{0.98\linewidth}{!}{
\begin{tabular}{l|c c | c c | c c}
\multirow{2}{*}{Source}   &\multicolumn{2}{ c| }{$\sigma_{\pp}$ [\%]}  &\multicolumn{2}{ c }{$N_\mathrm{AA}$ [\%]}  &\multicolumn{2}{ |c }{\RAA[\%]}\\
&$c\rightarrow\mu$  &$b\rightarrow\mu$  &$c\rightarrow\mu$  &$b\rightarrow\mu$  &$c\rightarrow\mu$  &$b\rightarrow\mu$\\
\hline
Muon efficiency			    & 0.5--1.0            & 0.4--0.6          & 0.6--16~           & 0.3--16~	        & 0.3--16~           & 0.2--16~  \\
Background removal		    & 4.3--12~           & 0.8--3.8          & 2.5--30~           & 1.0--5.1	        & 1.9--27~           & 0.5--4.7  \\
Charm--bottom separation	    & 4.5--9.8            & 3.2--8.0          & 9.2--37~           & 6.2--16~	        & 5.1--23~           & 4.1--13~  \\
Global normalization	            & $1.6$                & $1.6$	       & 0.9--4.6            & 0.9--4.6                & 1.8--4.9            & 1.8--4.9  \\
\hline
Total systematic uncertainty & 5.8--13~            & 3.1--7.4          & 10--47              & 6.9--19~               & 6.5--35~           & 5.4--18~ \\
\end{tabular}
}
\end{center}
\end{table}

\section{Results}

\begin{figure}[hbt]
\centering
\includegraphics[width=0.4\linewidth]{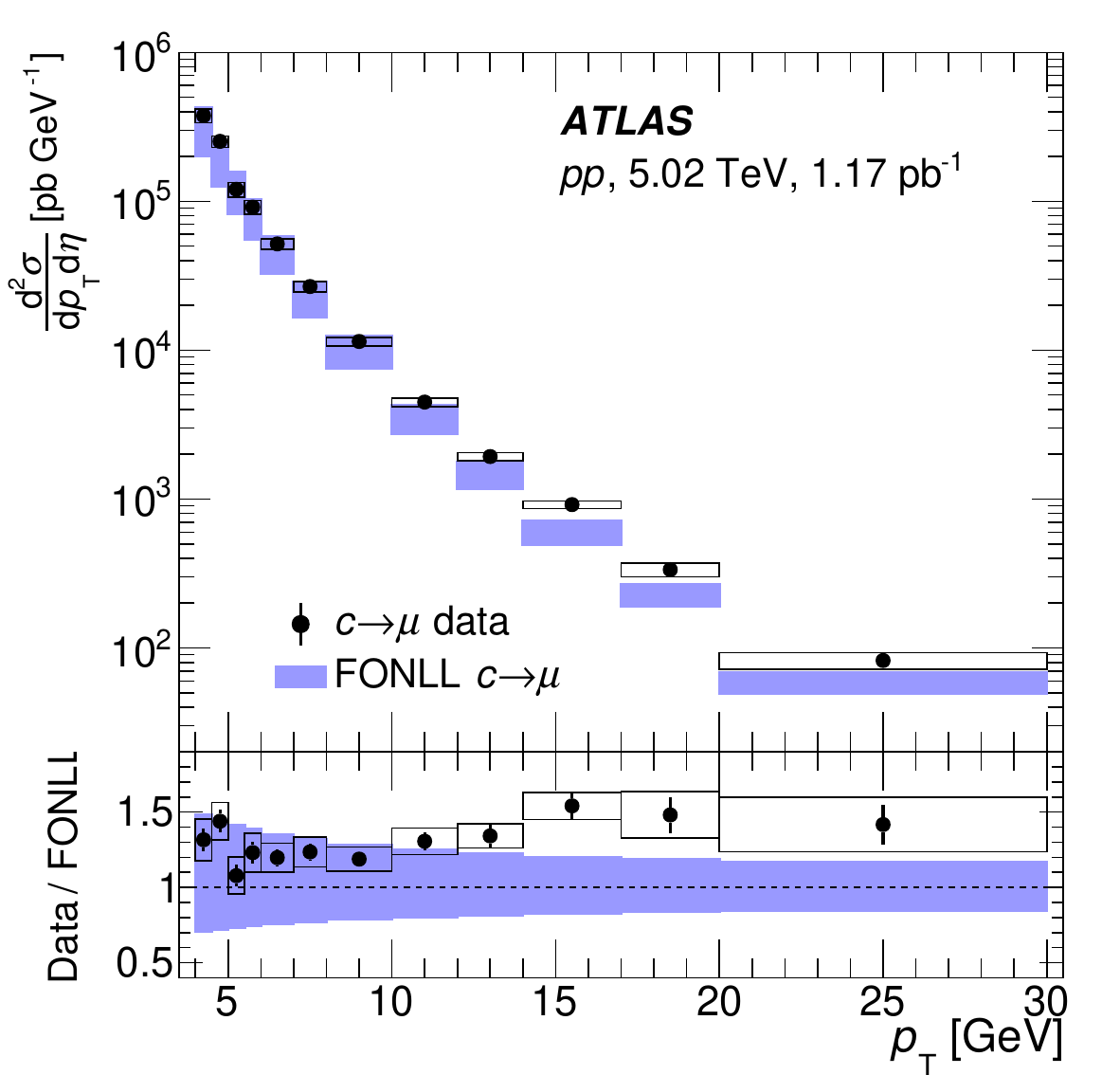}
\includegraphics[width=0.4\linewidth]{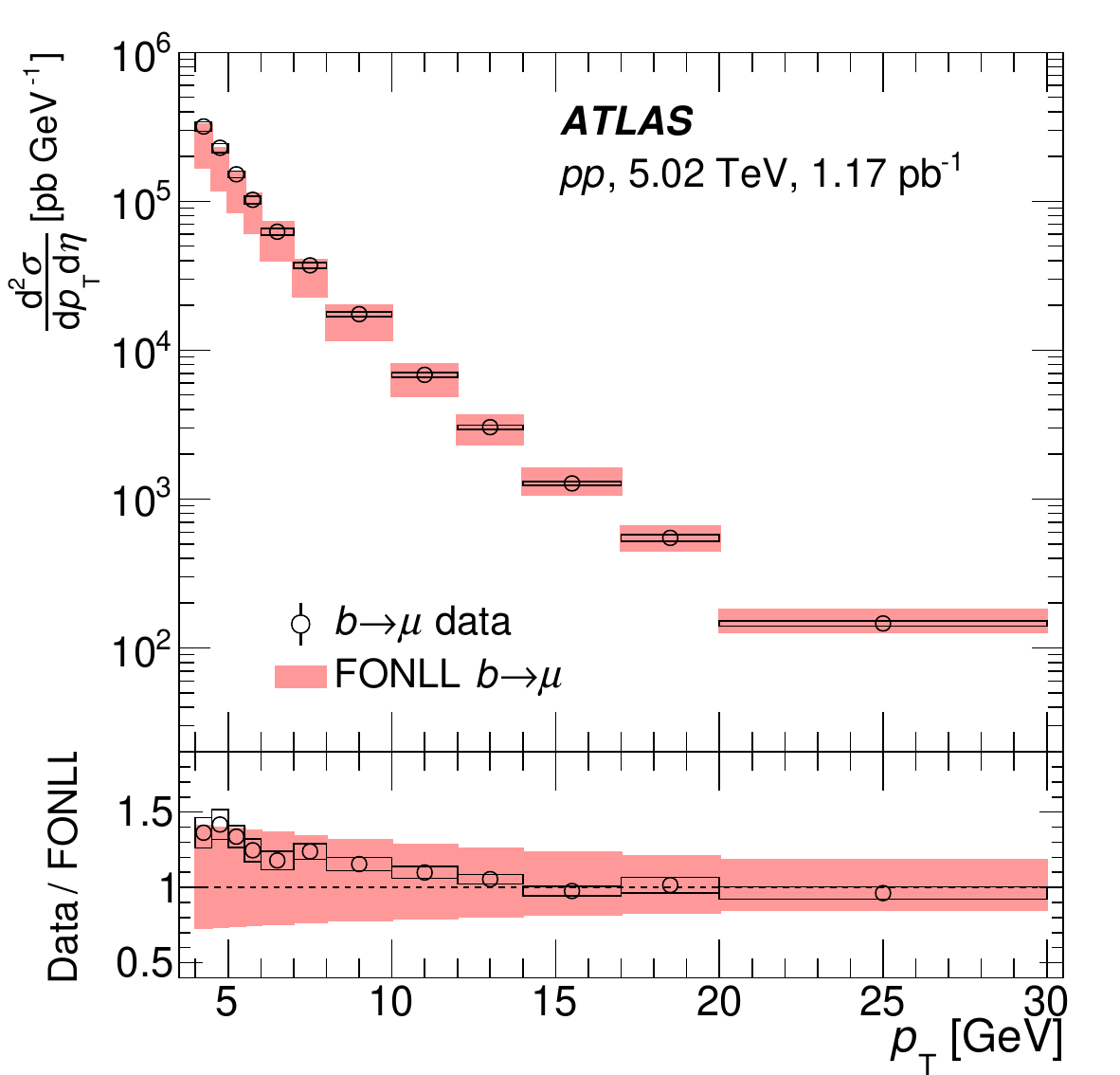}
\caption{\label{fig:result_ppXsec} Differential cross section of charm (left) and bottom (right) muons as a function of muon \pt, plotted at the centres of the \pt intervals, in \pp collisions at $\sqrt{s} = 5.02~\TeV$ in comparison with FONLL calculations.
Data results and FONLL calculations for bottom muons both include the $b\rightarrow c\rightarrow \mu$ contribution.
Statistical uncertainties in the data are shown as vertical lines and systematic uncertainties in the data and calculation are shown as boxes.
}
\end{figure}

\Fig{\ref{fig:result_ppXsec}} shows the differential cross section for muons from charm and bottom hadron decays within $\left|\eta\right| < 2$ as a function of muon \pT in \pp collisions at $\sqrt{s} = 5.02~\TeV$.  The measurements are compared with theoretical calculations for muons from $D$ and $B$ meson decays in the
FONLL resummation framework~\cite{Cacciari:2012ny}.
Uncertainties affecting the FONLL calculations include uncertainties in the parton distribution functions, heavy-flavour quark masses, and the renormalization and factorization scales.
For muons from charm quarks,
the FONLL calculation reaches the experimental data with the upper edge of its uncertainty band at $\pt < 10~\GeV$,
but underestimates the data at higher \pt.
Similar differences between FONLL calculations and prompt-charm measurements were observed in previous measurements at LHC energies,
for example, in ALICE measurements of prompt $D$ mesons~\cite{Acharya:2019mgn} and inclusive heavy-flavour leptons~\cite{Acharya:2019mky, Acharya:2019mom}, and in an LHCb measurement of prompt $D$ mesons~\cite{Aaij:2016jht}.
The measured cross section for muons from bottom quarks is a factor of 1.3--1.4 higher than the FONLL-calculated central value (including $b\rightarrow c\rightarrow \mu$) but still inside the FONLL calculation's uncertainty band at low \pt, while the calculated central value agrees with the data within experimental uncertainties for $\pt > 10~\GeV$, similar to observations in a previous ATLAS measurement of non-prompt charmonium~\cite{BPHY-2012-02} and an ALICE measurement of non-prompt $D$ mesons~\cite{Acharya:2021cqv}.

\begin{figure}[hbt]
\centering
\includegraphics[width=0.4\linewidth]{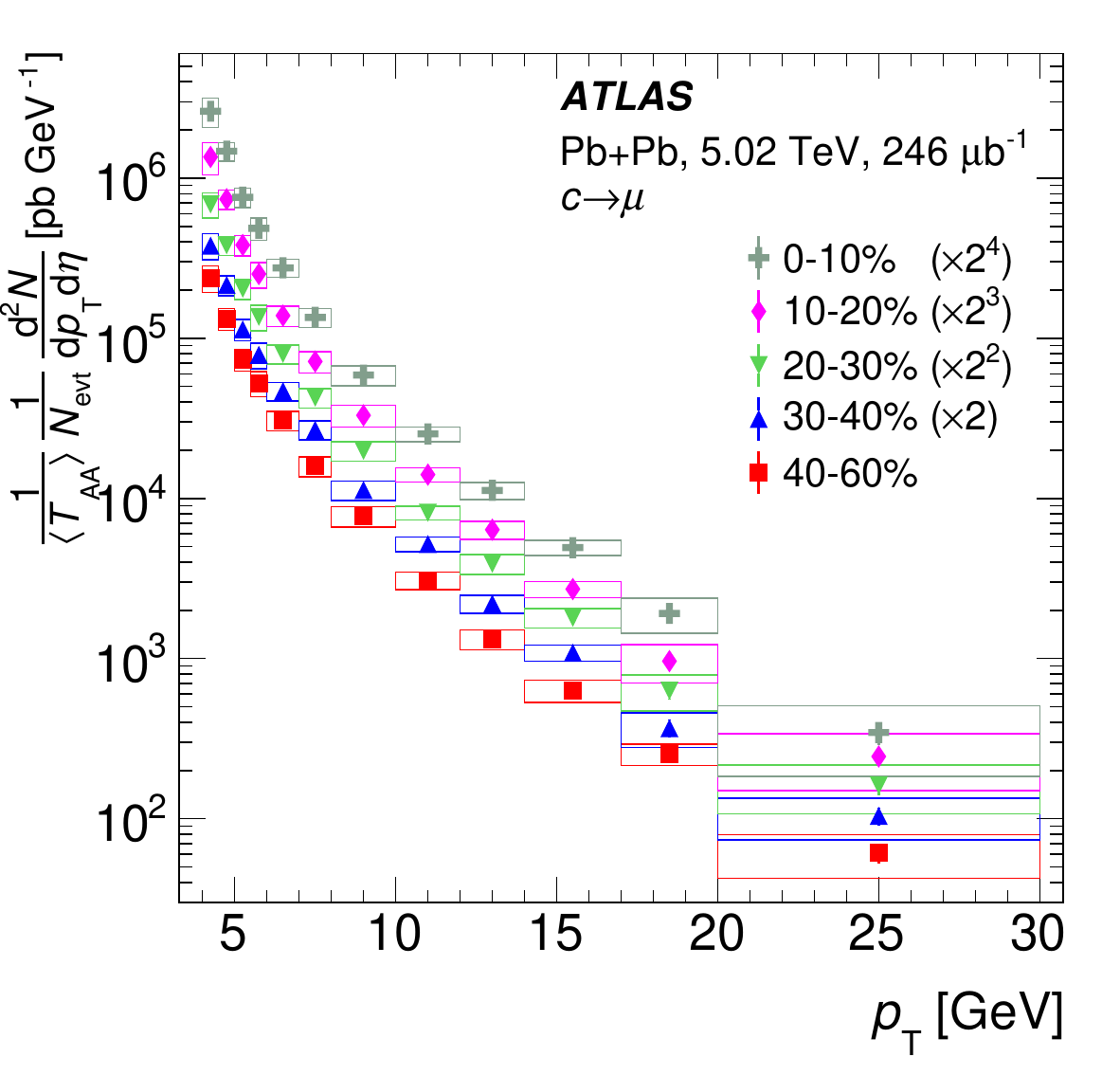}
\includegraphics[width=0.4\linewidth]{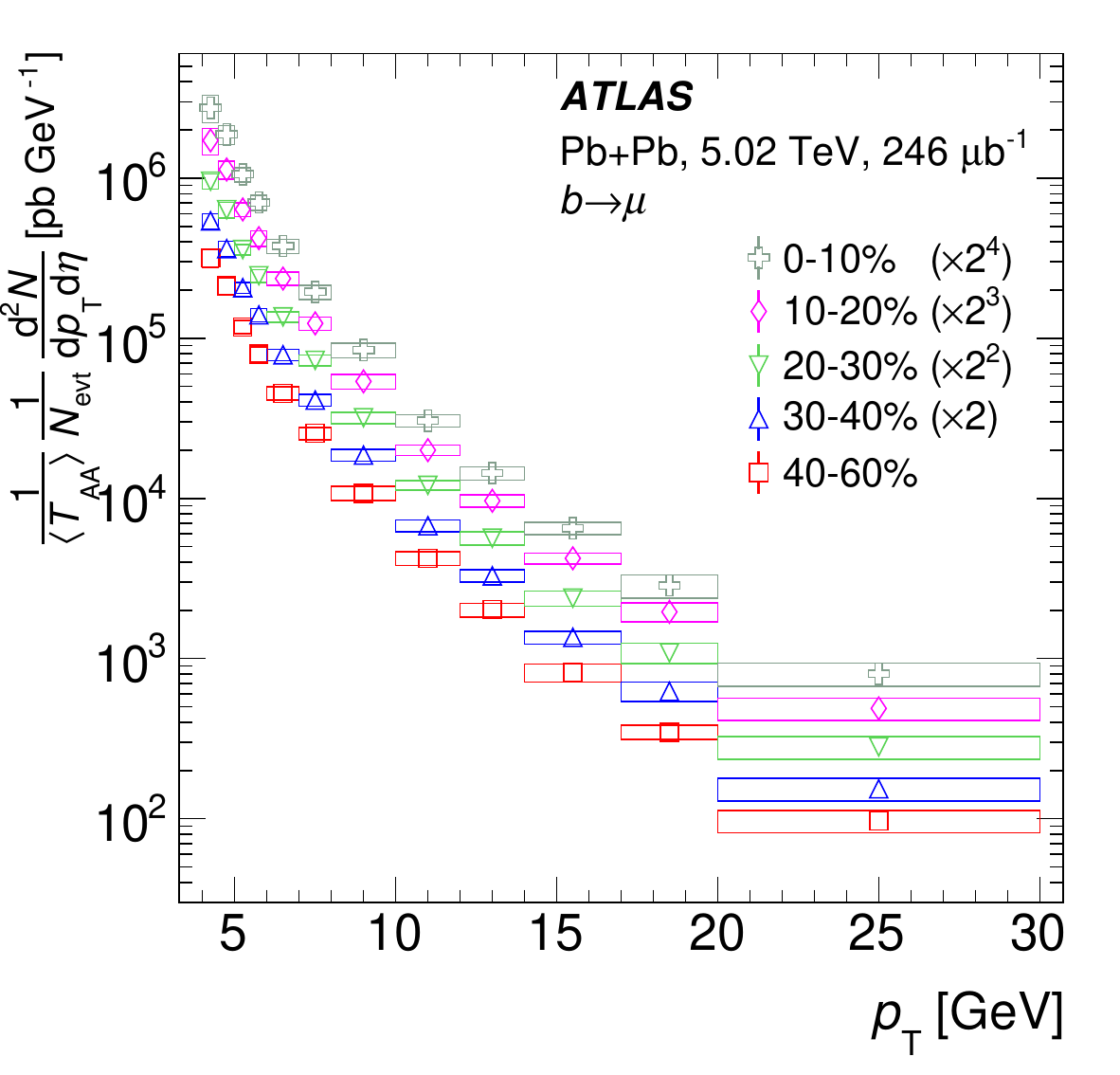}
\caption{\label{fig:result_PbPbYield} Differential per-event invariant yields of muons from charm hadron decays (left) and bottom hadron decays (right) as a function of \pt, plotted at the centres of the \pt intervals, for different centrality intervals in \PbPb collisions at $\sqn = 5.02~\TeV$.
For each centrality interval from peripheral to central, an additional scaling factor of $2$ is applied to the plotted points for visual clarity.
Statistical uncertainties are shown as vertical lines and systematic uncertainties as boxes.
}
\end{figure}
 
\begin{figure}[thb]
\centering
\includegraphics[width=0.4\linewidth]{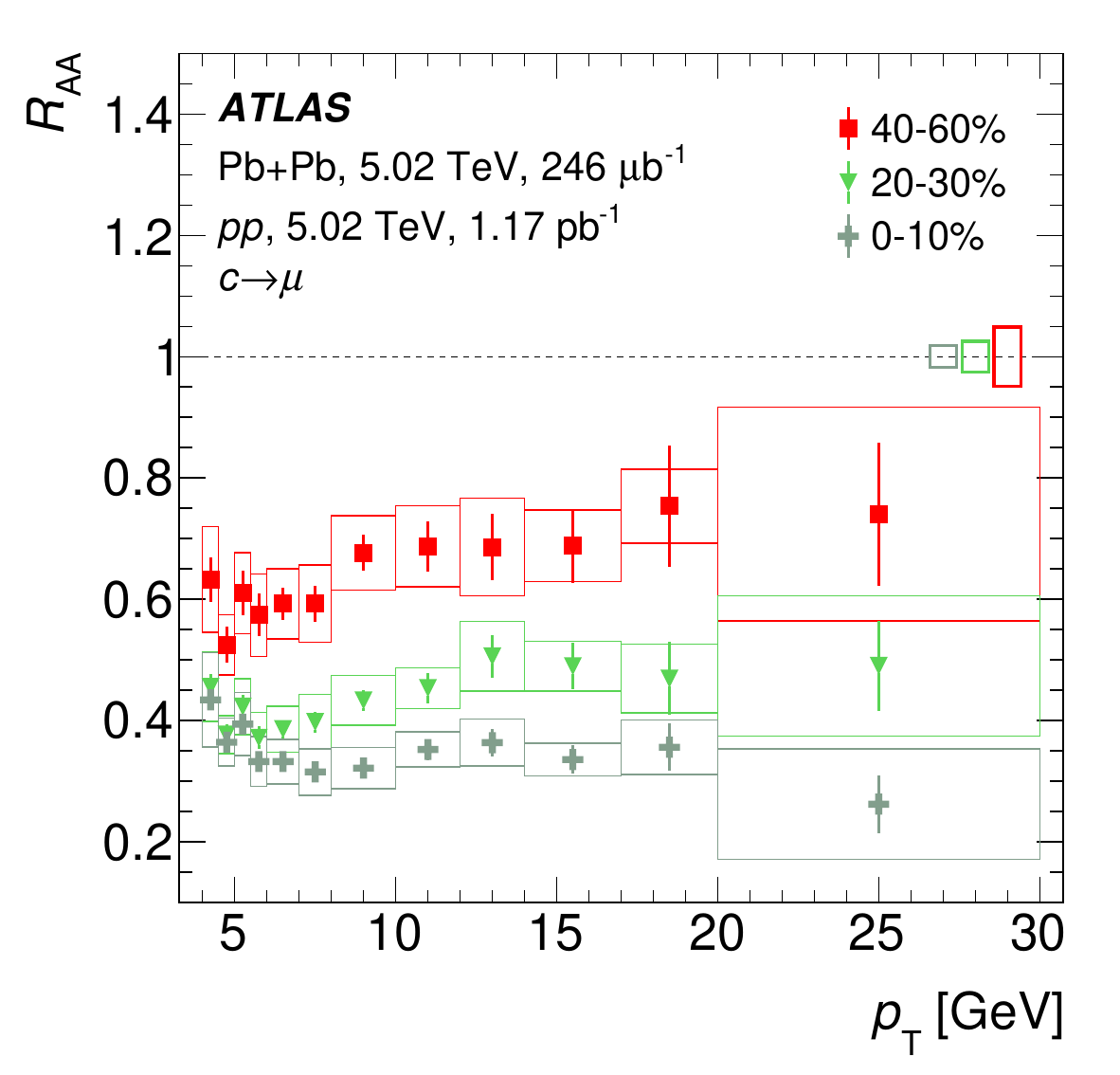}
\includegraphics[width=0.4\linewidth]{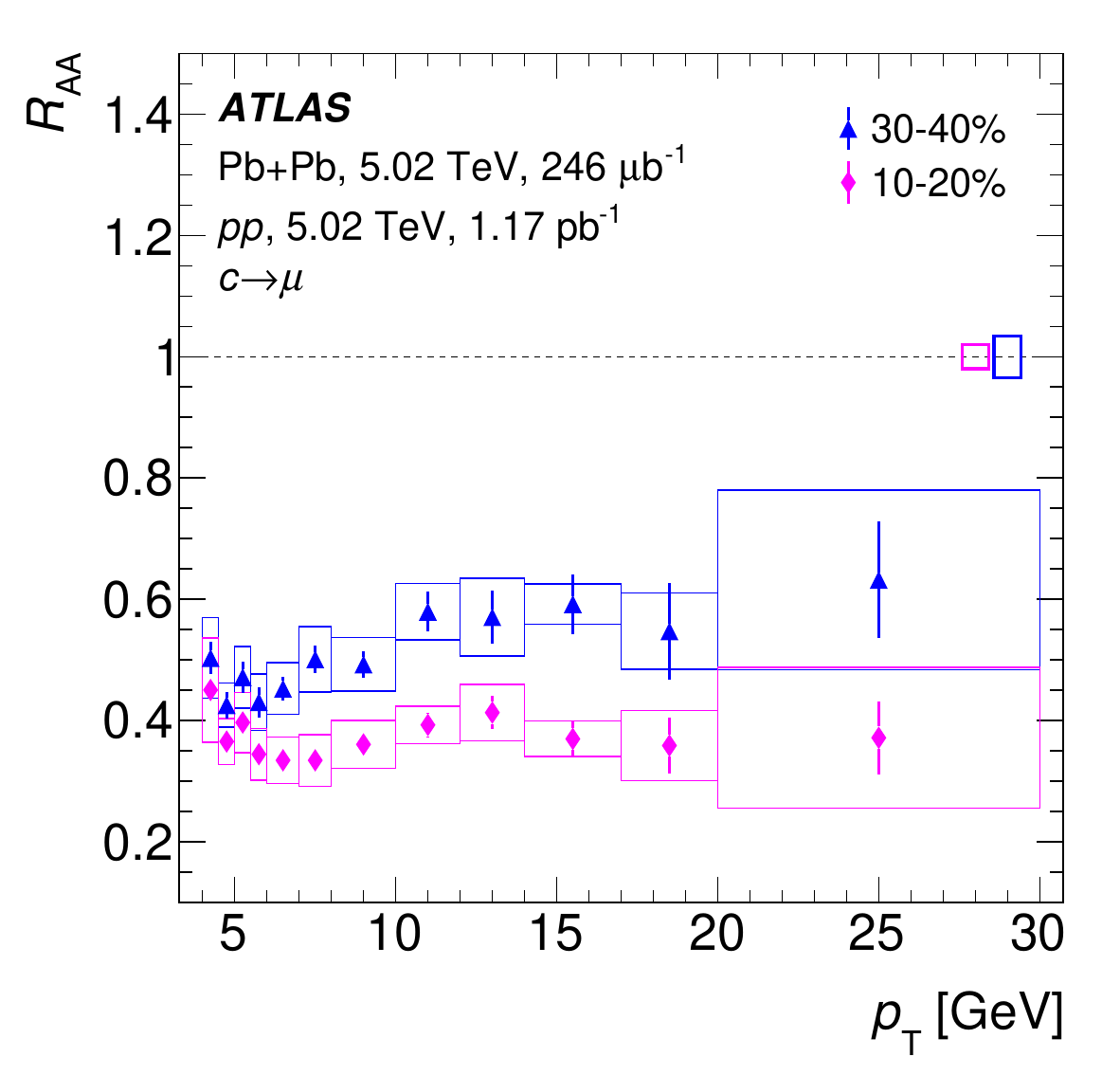}\\
\includegraphics[width=0.4\linewidth]{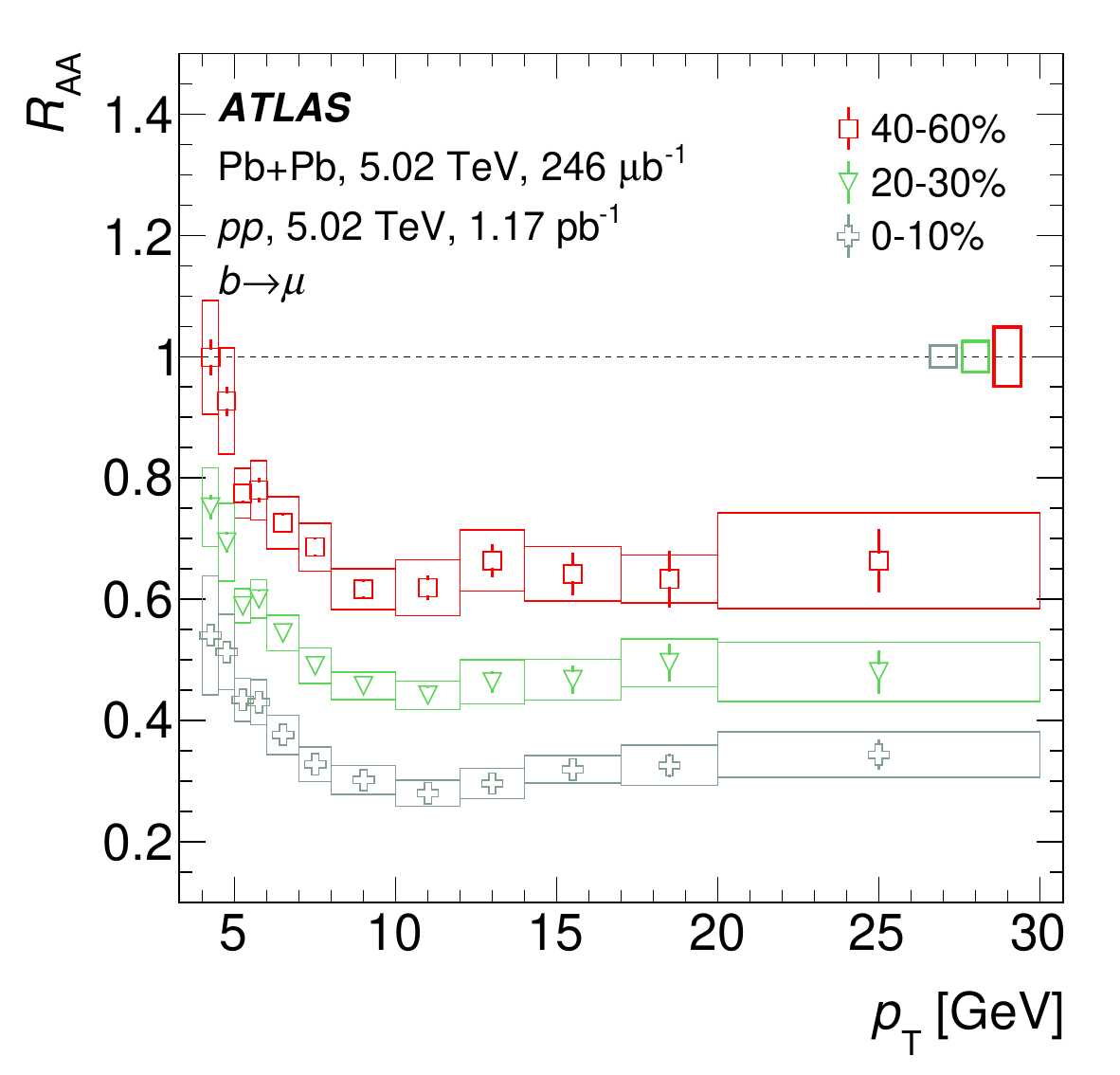}
\includegraphics[width=0.4\linewidth]{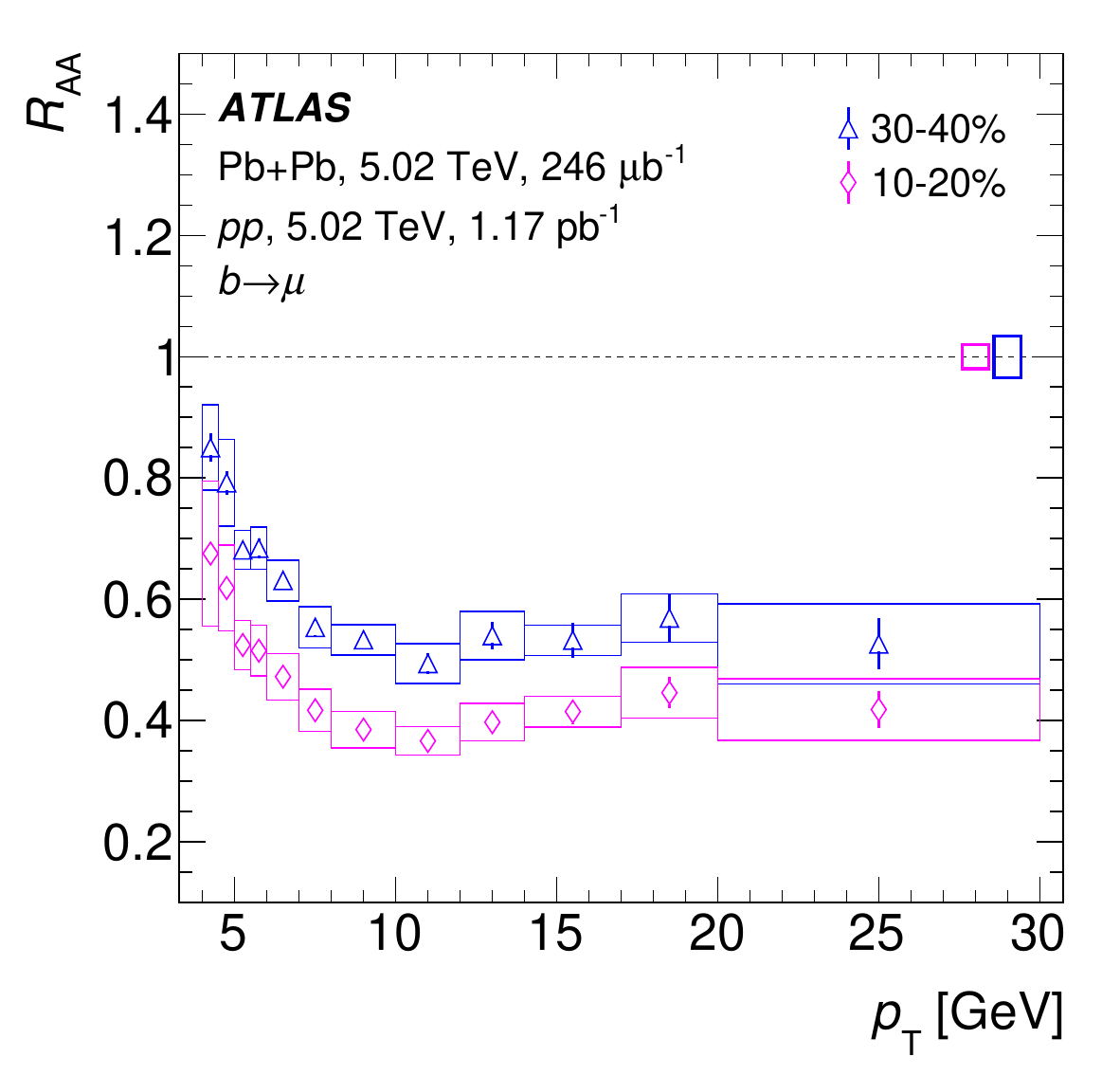}\\
\caption{\label{fig:result_raa} Nuclear modification factor, \RAA, for muons from charm hadron decays (top panels) and bottom hadron decays (bottom panels) as a function of \pt for five different centrality intervals.   The centrality intervals are separated for clarity into the left (0--10\%, 20--30\%, 40--60\%) and right (10--20\%, 30--40\%) panels.
Statistical uncertainties are shown as vertical lines and uncorrelated systematic uncertainties as boxes around the points. Correlated fractional systematic uncertainties, including \TAA and \pp luminosity uncertainties, are isolated as coloured boxes around unity for different centrality intervals.
}
\end{figure}

\Fig{\ref{fig:result_PbPbYield}} shows the differential per-event invariant yields for muons from charm and bottom hadron decays, in \PbPb collisions at $\sqn = 5.02~\TeV$ as a function of muon \pt.
To quantify the modification of the momentum distribution between \pp and \PbPb collisions
the nuclear modification factor is calculated according to Eq.~(\ref{eqn:raa}) in several centrality intervals.
The resulting charm and bottom muon \RAA values are shown in \Fig{\ref{fig:result_raa}} as a function of muon \pt.
There is substantial suppression of muons from both charm and bottom hadron decays for all analysed \PbPb centrality intervals.  The suppression increases monotonically from the 40--60\% interval to the most central 0--10\% interval.
The monotonic centrality dependence follows expectations as the heavy quarks spend a longer time in the hotter and larger QGP droplet formed in collisions that have larger nuclear overlap, i.e.\ the more central intervals.
Charm muon \RAA shows weak \pt dependence
in all centrality intervals, while bottom muon \RAA first decreases with increasing muon \pt up to $10~\GeV$ and then remains mostly unchanged in all centralities.

\begin{figure}[hbt]
\centering
\includegraphics[width=0.45\linewidth]{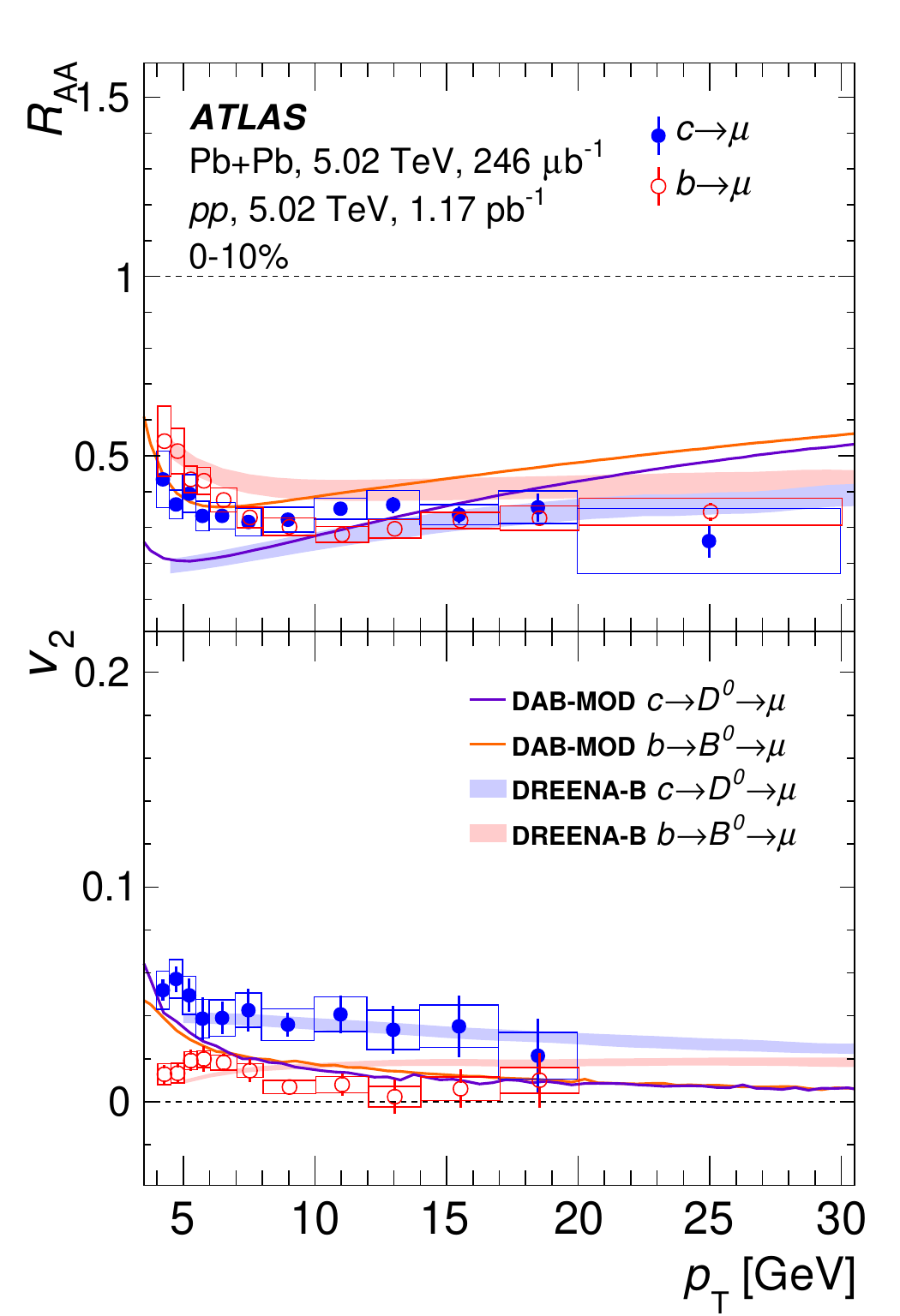}
\includegraphics[width=0.45\linewidth]{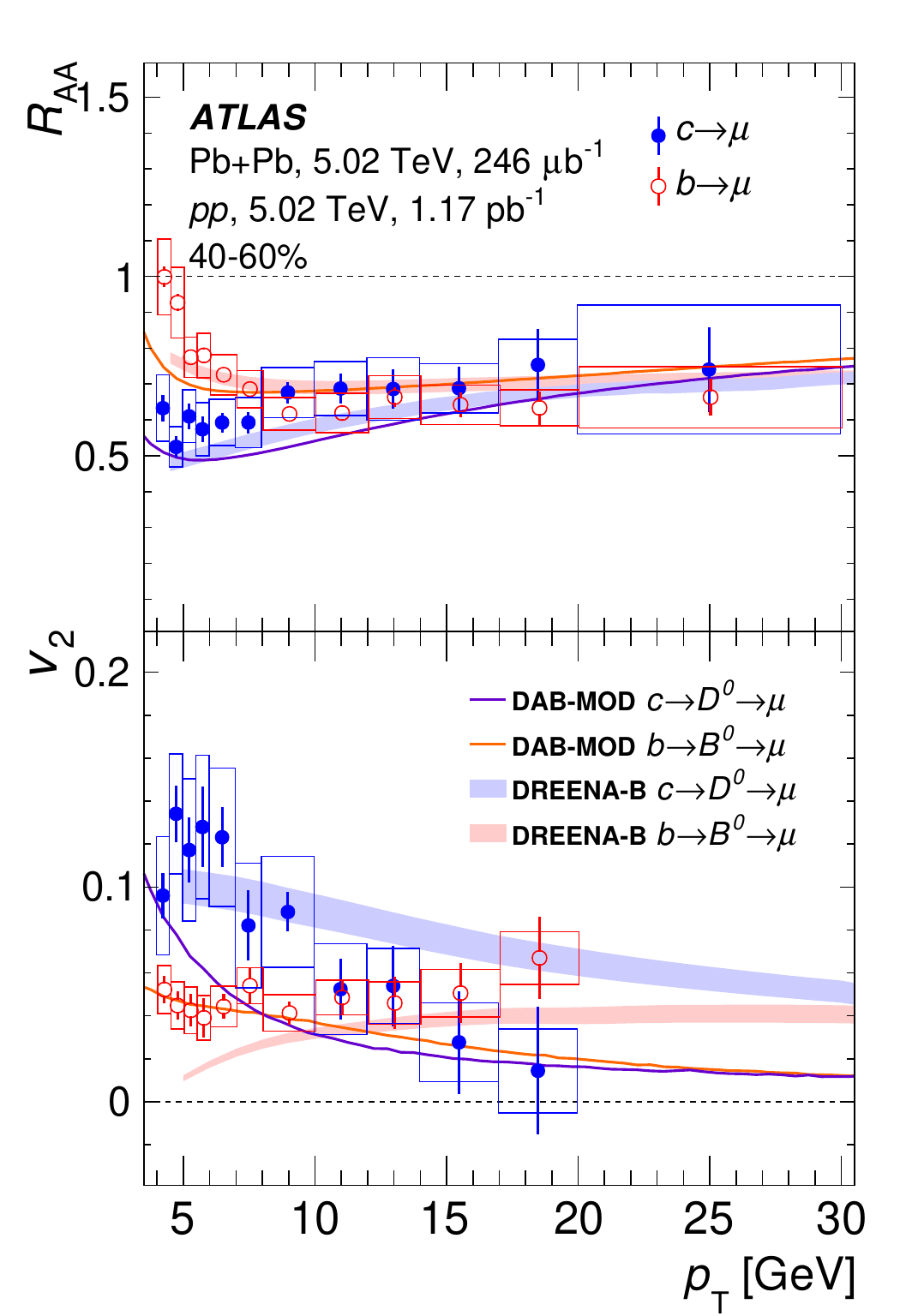}
\caption{\label{fig:result_raa_vsCalc} Nuclear modification factor, \RAA, (top) and elliptic flow, $v_2$, results taken from Ref.~\cite{HION-2019-11} (bottom) for muons from bottom hadron decays and charm hadron decays for 0--10\% (left) and 40--60\% (right) centrality intervals as a function of \pt.
Statistical uncertainties are shown as vertical lines and systematic uncertainties as boxes.
Also shown are theoretical calculations from the \textsc{dreena-b} and \textsc{dab-mod} models.
See text regarding the uncertainty band of \textsc{dreena-b} model calculations.
}
\end{figure}

Muons from charm decays have a stronger suppression than muons from bottom decays at low \pt in all centrality intervals.
This difference in suppression is highlighted in \Fig{\ref{fig:result_raa_vsCalc}}, which overlays the 0--10\% and 40--60\% centrality intervals already presented in \Fig{\ref{fig:result_raa}} for both charm and bottom muon \RAA.
The mass ordering of the measured \RAA follows expectations, as the charm quarks, being lighter than the bottom quarks, are expected to lose more energy in the QGP.
Thus the charm quarks are pushed further to lower \pT, and emerge more strongly suppressed than bottom quarks when compared with the cross section in \pp collisions.
The decay muon \pT spectrum depends on the \pT spectra of various charm and bottom hadrons in pp and Pb+Pb collisions. A portion of the difference found in the charm and bottom muon \RAA results could be attributed to the difference between charm and bottom hadron spectra, and thus full model comparisons are necessary.
 
\Fig{\ref{fig:result_raa_vsCalc}} also compares the experimental data with theoretical calculations referred to as \textsc{dreena-b}~\cite{Zigic:2018ovr} and \textsc{dab-mod}~\cite{Katz:2019fkc,Prado:2016szr}.
The \textsc{dreena-b} calculation includes radiative and collisional energy loss of the heavy quarks traversing the QGP, the latter modelled via a 1~+~1D Bjorken expansion~\cite{PhysRevD.27.140} with path-length distributions calculated following the
procedure described in Ref.~\cite{Dainese:2003wq}.
The width of the band corresponding to the \textsc{dreena-b} theoretical uncertainties reflects the range of the ratio of magnetic to electric screening masses as constrained by non-perturbative calculations~\cite{PhysRevD.27.140}.
As discussed in Ref.~\cite{Zigic:2018ovr}, the predicted \RAA is higher for $B$ mesons than for $D$ mesons, converging to the same value at $\pt \approx 25~\GeV$ as is expected when the particle \pt becomes much larger than the mass of the heavier $b$-quark.
The corresponding \RAA for muons shown for direct comparison with the experimental data is calculated using \PYTHIA[8], which simulates meson decay kinematics.
The \textsc{dreena-b} prediction is in reasonable agreement with the experimental data.
The \textsc{dab-mod} framework used here includes calculations with only Langevin drag and diffusion contributions for the heavy quarks in the QGP.
The curves shown here are obtained with \textsc{Trento} geometric initial conditions~\cite{PhysRevC.92.011901}, heavy-quark Langevin dynamics with the Moore and Teaney parameterization~\cite{Moore:2004tg}, and coupling values for charm (bottom) of $D/2\pi T$ = 2.23 (2.79), where $D$ is the spatial diffusion coefficient and $T$ is the temperature.
The temperature at which heavy quarks decouple from the medium
is $T = 160~\MeV$ and both coalescence and fragmentation are implemented for hadronization.
The \textsc{dab-mod} predictions with only Langevin dynamics shown in \Fig{\ref{fig:result_raa_vsCalc}} as coloured solid lines are in qualitative agreement with the experimental data, but have a stronger \pt dependence than the experimental data, particularly in the most central collisions.
No uncertainties are included for \textsc{dab-mod} calculations shown here.

It is notable that both of these calculations also predict the azimuthal anisotropies of the heavy-flavour muons.
ATLAS has previously published azimuthal anisotropies
quantified by the elliptic flow coefficient, $v_2$, for muons from charm and bottom hadron decays~\cite{HION-2019-11}.
The lower panels of \Fig{\ref{fig:result_raa_vsCalc}} show those ATLAS measurements of $v_2$ as a function of \pt in comparison with both the
\textsc{dreena-b} and \textsc{dab-mod} calculations.
The \textsc{dreena-b} calculations agree qualitatively with both $v_2$ and \RAA for both charm and bottom muons, while the previously described implementation of \textsc{dab-mod} underestimates the charm muon $v_2$ even though it qualitatively matches the \RAA.
In all theoretical implementations, a larger coupling of charm and bottom quarks to the QGP results in a reduced \RAA (i.e.\ more suppression) and an increased $v_2$ (i.e.\ larger anisotropy).    Thus, increasing the coupling of charm to the QGP in \textsc{dab-mod}, for example, could
bring the predicted $v_2$ into
closer agreement with data but would simultaneously decrease \RAA, pushing the calculation  further below the data.
That said, another key component of these calculations is the modelling of the QGP space-time evolution, and thus it could be instructive in the future to compare the different theory calculations with a common QGP model to test whether the differences in \RAA and $v_{2}$ arise from the QGP modelling or the energy-loss implementation.

\begin{figure}[hbt]
\centering
\includegraphics[width=0.4\linewidth]{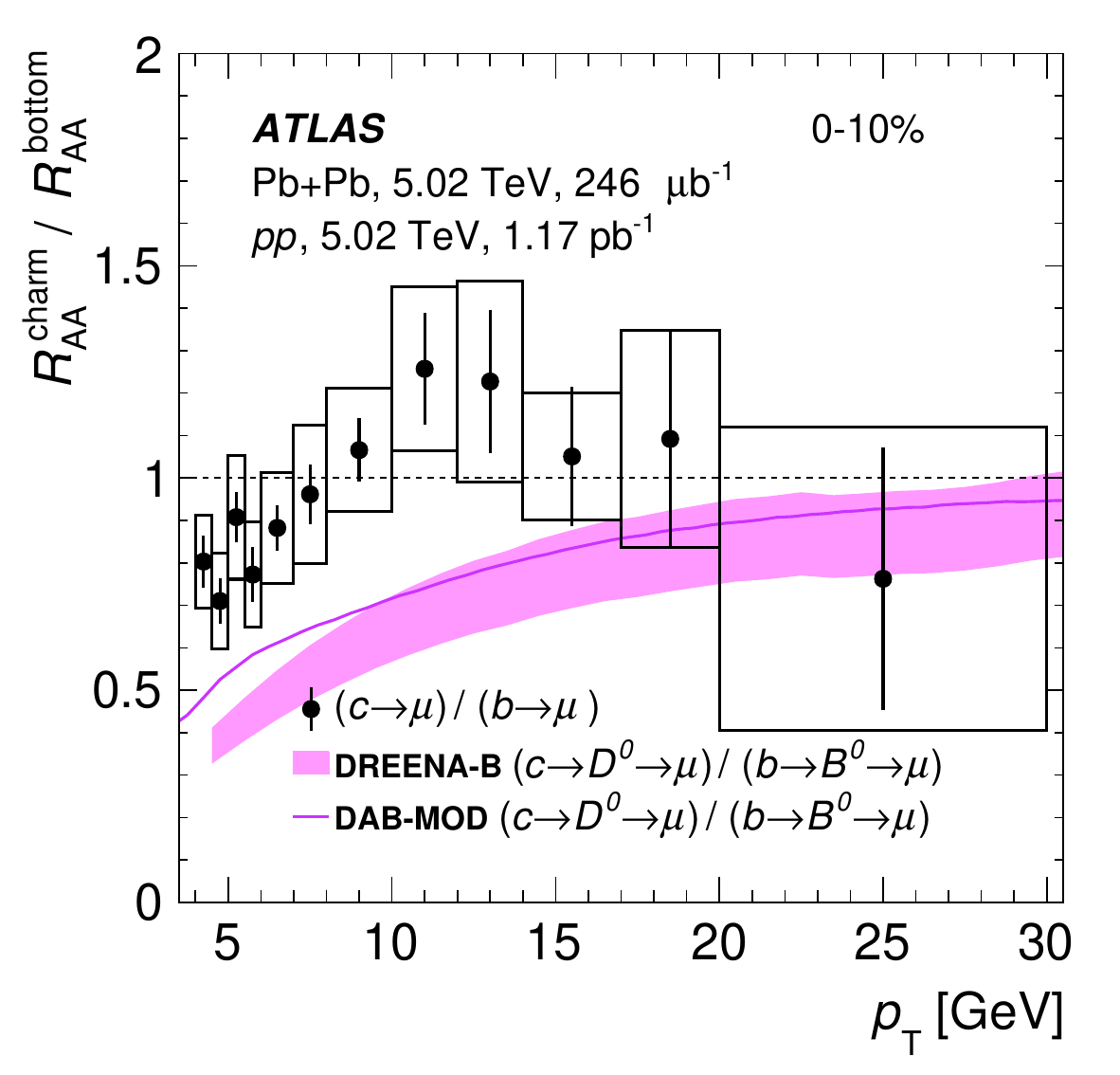}
\includegraphics[width=0.4\linewidth]{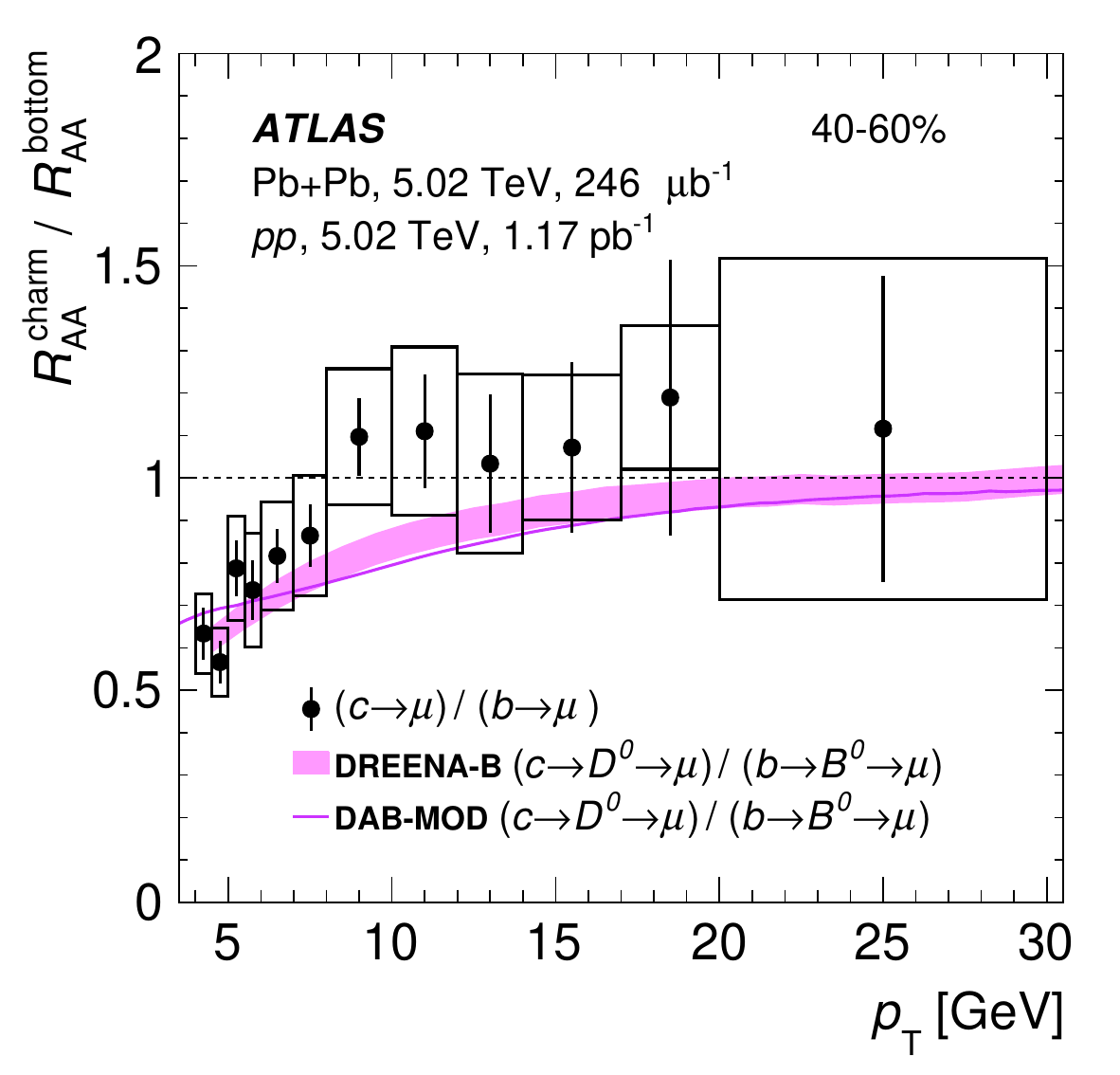}
\caption{\label{fig:result_raa_vsCalcRatio} The ratio of charm muon \RAA to bottom muon \RAA, $\RAA^\mathrm{charm}/\RAA^\mathrm{bottom}$, for 0--10\% (left) and 40--60\% (right) \PbPb centrality intervals as a function of muon \pt.
Statistical uncertainties are shown as vertical lines and systematic uncertainties as boxes.
Also shown are theoretical calculations for the decay muons from \textsc{dreena-b} and \textsc{dab-mod} in the same centrality intervals.
}
\end{figure}

\Fig{\ref{fig:result_raa_vsCalcRatio}} shows the measured ratio of charm muon \RAA to bottom muon \RAA as a function of \pt in comparison with \textsc{dreena-b} and \textsc{dab-mod} calculations.
The large uncertainty in the measured ratio is due to a strong negative correlation between the charm and bottom muon \RAA uncertainties.
As indicated by the measured \RAA ratios, charm muons are significantly more suppressed than bottom muons in the $\pt < 8~\GeV$ range.
However, no strong conclusion about the relative strength of their suppression can be drawn at higher \pt with existing large uncertainties.
Compared with the ratios measured in data, the calculations underestimate the \RAA ratios for 0--10\% centrality,
while they mostly capture the magnitude and \pt dependence of the ratio for 40--60\% centrality.
As shown in \Fig{\ref{fig:result_raa_vsCalc}}, the discrepancy between data and the models in the 0--10\% centrality interval is primarily due to the underestimation of charm muon \RAA.

\clearpage
\section{Conclusion}
 
The ATLAS experiment at the LHC has measured the production rates and nuclear modification factors, \RAA,  of muons from semileptonic decays of heavy-flavour hadrons in \pp and \PbPb collisions at 5.02~\TeV.
The measurement uses 2017 \pp data and combined 2015 and 2018 \PbPb data corresponding to integrated luminosities of $1.17~\ipb$ and $246~\mathrm{\mu b}^{-1}$ respectively.
Compared to a previous ATLAS measurement of heavy-flavour muons, this Letter reports separate results for charm and bottom muons
and covers a wider \pt range.
The differential cross section measured in \pp collisions for muons from decays of hadrons containing a bottom quark is reproduced with FONLL calculations.
For muons from decays of hadrons containing a charm quark, the FONLL calculation is lower than the data at all measured \pt but agrees with it within the systematic uncertainties of the calculations below $10~\GeV$.
The \RAA measurements indicate a significant suppression of the yield of muons from both charm and bottom hadron decays, with a suppression that increases monotonically from peripheral to central collisions.
 
The suppression is stronger for charm quarks than for bottom quarks, as seen via muons at low $\pT < 10~\GeV$.  The relative difference between charm and bottom suppression is consistent with theoretical expectations for 10-60\% \PbPb collisions, while the difference is smaller in data compared to theory for 0-10\% \PbPb collisions.   The simultaneous constraints imposed by the measurements of \RAA presented here and the flow measurements previously published by ATLAS could provide important information for understanding heavy-quark transport and QGP properties.

\section*{Acknowledgements}


We thank CERN for the very successful operation of the LHC, as well as the
support staff from our institutions without whom ATLAS could not be
operated efficiently.
 
We acknowledge the support of
ANPCyT, Argentina;
YerPhI, Armenia;
ARC, Australia;
BMWFW and FWF, Austria;
ANAS, Azerbaijan;
SSTC, Belarus;
CNPq and FAPESP, Brazil;
NSERC, NRC and CFI, Canada;
CERN;
ANID, Chile;
CAS, MOST and NSFC, China;
Minciencias, Colombia;
MEYS CR, Czech Republic;
DNRF and DNSRC, Denmark;
IN2P3-CNRS and CEA-DRF/IRFU, France;
SRNSFG, Georgia;
BMBF, HGF and MPG, Germany;
GSRI, Greece;
RGC and Hong Kong SAR, China;
ISF and Benoziyo Center, Israel;
INFN, Italy;
MEXT and JSPS, Japan;
CNRST, Morocco;
NWO, Netherlands;
RCN, Norway;
MEiN, Poland;
FCT, Portugal;
MNE/IFA, Romania;
JINR;
MES of Russia and NRC KI, Russian Federation;
MESTD, Serbia;
MSSR, Slovakia;
ARRS and MIZ\v{S}, Slovenia;
DSI/NRF, South Africa;
MICINN, Spain;
SRC and Wallenberg Foundation, Sweden;
SERI, SNSF and Cantons of Bern and Geneva, Switzerland;
MOST, Taiwan;
TAEK, Turkey;
STFC, United Kingdom;
DOE and NSF, United States of America.
In addition, individual groups and members have received support from
BCKDF, CANARIE, Compute Canada and CRC, Canada;
COST, ERC, ERDF, Horizon 2020 and Marie Sk{\l}odowska-Curie Actions, European Union;
Investissements d'Avenir Labex, Investissements d'Avenir Idex and ANR, France;
DFG and AvH Foundation, Germany;
Herakleitos, Thales and Aristeia programmes co-financed by EU-ESF and the Greek NSRF, Greece;
BSF-NSF and GIF, Israel;
Norwegian Financial Mechanism 2014-2021, Norway;
NCN and NAWA, Poland;
La Caixa Banking Foundation, CERCA Programme Generalitat de Catalunya and PROMETEO and GenT Programmes Generalitat Valenciana, Spain;
G\"{o}ran Gustafssons Stiftelse, Sweden;
The Royal Society and Leverhulme Trust, United Kingdom.
 
The crucial computing support from all WLCG partners is acknowledged gratefully, in particular from CERN, the ATLAS Tier-1 facilities at TRIUMF (Canada), NDGF (Denmark, Norway, Sweden), CC-IN2P3 (France), KIT/GridKA (Germany), INFN-CNAF (Italy), NL-T1 (Netherlands), PIC (Spain), ASGC (Taiwan), RAL (UK) and BNL (USA), the Tier-2 facilities worldwide and large non-WLCG resource providers. Major contributors of computing resources are listed in Ref.~\cite{ATL-SOFT-PUB-2021-003}.
 

\printbibliography

\clearpage 
 
\begin{flushleft}
\hypersetup{urlcolor=black}
{\Large The ATLAS Collaboration}

\bigskip

\AtlasOrcid[0000-0002-6665-4934]{G.~Aad}$^\textrm{\scriptsize 99}$,    
\AtlasOrcid[0000-0002-5888-2734]{B.~Abbott}$^\textrm{\scriptsize 124}$,    
\AtlasOrcid[0000-0002-7248-3203]{D.C.~Abbott}$^\textrm{\scriptsize 100}$,    
\AtlasOrcid[0000-0002-2788-3822]{A.~Abed~Abud}$^\textrm{\scriptsize 34}$,    
\AtlasOrcid[0000-0002-1002-1652]{K.~Abeling}$^\textrm{\scriptsize 51}$,    
\AtlasOrcid[0000-0002-2987-4006]{D.K.~Abhayasinghe}$^\textrm{\scriptsize 91}$,    
\AtlasOrcid[0000-0002-8496-9294]{S.H.~Abidi}$^\textrm{\scriptsize 27}$,    
\AtlasOrcid[0000-0001-5329-6640]{H.~Abramowicz}$^\textrm{\scriptsize 157}$,    
\AtlasOrcid[0000-0002-1599-2896]{H.~Abreu}$^\textrm{\scriptsize 156}$,    
\AtlasOrcid[0000-0003-0403-3697]{Y.~Abulaiti}$^\textrm{\scriptsize 5}$
% ...
\bigskip
\\

% ...
$^{5}$High Energy Physics Division, Argonne National Laboratory, Argonne IL; United States of America.\\
% ...
$^{27}$Physics Department, Brookhaven National Laboratory, Upton NY; United States of America.\\
% ...
$^{34}$CERN, Geneva; Switzerland.\\
% ...
$^{51}$II. Physikalisches Institut, Georg-August-Universit\"{a}t G\"ottingen, G\"ottingen; Germany.\\
% ...
$^{91}$Department of Physics, Royal Holloway University of London, Egham; United Kingdom.\\
% ...
$^{99}$CPPM, Aix-Marseille Universit\'e, CNRS/IN2P3, Marseille; France.\\
$^{100}$Department of Physics, University of Massachusetts, Amherst MA; United States of America.\\
% ...
$^{124}$Homer L. Dodge Department of Physics and Astronomy, University of Oklahoma, Norman OK; United States of America.\\
% ...
$^{156}$Department of Physics, Technion, Israel Institute of Technology, Haifa; Israel.\\
$^{157}$Raymond and Beverly Sackler School of Physics and Astronomy, Tel Aviv University, Tel Aviv; Israel.\\
% ...

\end{flushleft}


\end{document}